\documentclass{article}

\usepackage{graphicx}
\usepackage{amssymb}
\usepackage{comment}
\usepackage{color}
\usepackage{enumerate}
\usepackage{subfigure}
\usepackage{fullpage}
\usepackage{multirow}
\usepackage{epsfig}
\usepackage[ruled, noend, lined]{algorithm2e}

\newtheorem{theorem}{Theorem}
\newtheorem{lemma}[theorem]{Lemma}
\newtheorem{corollary}[theorem]{Corollary}

\newcommand{\buzz}[1]{\textbf{#1}}
\newcommand{\op}[1]{\texttt{#1}}
\newcommand{\msg}[2]{$\langle$\op{#1}, #2$\rangle$}
\newcommand{\msgnop}[1]{$\langle$\op{#1}$\rangle$}

\newcommand{\restrictedto}{\upharpoonright}
\newcommand{\prefix}{\preceq}
\newcommand{\nil}{\bot}

\newcommand{\newloglike}[2]{\newcommand{#1}{\mathop{\rm #2}\nolimits}}
\newloglike{\E}{E}
\newcommand{\LemmaDeltaZeroBound}{\sum_{r=1}^{s-1} {m+1 \choose r}{n-m-1 \choose r-1}}
\newcommand{\LemmaDeltaZeroBoundOne}{\sum_{r=1}^{s_1-1} {m+1 \choose r}{n-m-1 \choose r-1}}
\newcommand{\LemmaDeltaHFloorThing}{\left\lfloor\frac{2}{3}\cdot 2^{\ell} \right\rfloor}
\newcommand{\LemmaDeltaHLHS}{\Pr\left[|\delta_{\ell} A| \le \frac{1}{3}\cdot 2^{\ell} \right]}
\newcommand{\etal}[1]{{\it et al.\/}}

\newcommand{\neigh}[3]{$#1.$neighbor[$#2$][$#3$]}

\dontprintsemicolon
\SetKw{KwAlgorithm}{Algorithm}
\SetKw{KwSend}{send}
\SetKw{KwGoTo}{goto}
\SetKw{KwBreak}{break}
\SetKw{KwDownTo}{downto}
\decmargin{2em}

\includecomment{full}
\excludecomment{abs}

\excludecomment{suppress}

\begin{full}
\newenvironment{proof}{\noindent\par{\bf Proof: }}{\nopagebreak\rule{1 ex}{0.8 em}\medskip}

\excludecomment{proofsketch}
\end{full}

\begin{document}

\title{Skip Graphs}
\author{
James Aspnes\thanks{
Department of Computer Science, Yale University,
New Haven, CT 06520-8285, USA.
Email: {\tt aspnes@cs.yale.edu}.
Supported by NSF grants CCR-9820888 and CCR-0098078.}
\and Gauri Shah\thanks{
Department of Computer Science, Yale University,
New Haven, CT 06520-8285, USA.
Email: {\tt shah@cs.yale.edu}.
Supported by NSF grants CCR-9820888 and CCR-0098078.}
}
\date{}

\maketitle


\begin{abstract}
Skip graphs are a novel distributed data structure, based on skip lists,
that provide the full functionality of a balanced tree in a distributed
system where resources are stored in separate nodes that may fail
at any time.
They are designed for use in searching peer-to-peer systems,
and by providing the ability to perform queries based on key ordering,
they improve on existing search tools that provide only hash table
functionality.
Unlike skip lists or other tree data structures,
skip graphs are highly resilient, tolerating a large fraction of
failed nodes without losing connectivity.
In addition, constructing, inserting new nodes into, searching
a skip graph, and detecting and repairing errors in the data
structure introduced by node failures can be done using simple and
straightforward algorithms.
\end{abstract}


\section{Introduction}
\label{section-introduction}

Peer-to-peer networks are distributed systems without any
central authority that are used for efficient location of
shared resources. Such systems have become very popular
for Internet applications in a short period of time.
A survey of recent peer-to-peer research yields a slew
of desirable features for a peer-to-peer systems
such as decentralization, scalability,
fault-tolerance, self-stabilization, data availability,
load balancing, dynamic addition and deletion of peer nodes,
efficient and complex query searching, incorporating
geography in searches and exploiting spatial as well as
temporal locality in searches.  The initial approaches,
such as those used by
Napster~\cite{Napster}, Gnutella~\cite{Gnutella} 
and Freenet~\cite{Freenet}, do not
support most of these features and are clearly unscalable
either due to the use of a central server (Napster) or due
to high message complexity from performing searches by
flooding the network (Gnutella).
The performance of Freenet is difficult to evaluate, but it provides
no provable guarantee on the search latency and permits accessible
data to be missed.

Recent peer-to-peer systems like CAN~\cite{RatnasamyFHKS2001},
Chord~\cite{StoicaMKKB2001}, Pastry~\cite{RowstronD2001},
Tapestry~\cite{ZhaoKJ2001} and Viceroy~\cite{MalkhiNR2002}
use a \buzz{distributed hash
table} (DHT) approach to overcome scalability problems. To
ensure scalability, they hash the key of a resource to
determine which node it will be stored at and
balance out the load on the nodes in the network. The
main operation in these networks is to retrieve the identity
of the node which stores the resource, from any
other node in the network. To this end, there is an overlay
graph in which the location of the nodes and resources is determined
by the hashed values of their identities and keys respectively.
Resource location using the overlay graph is done in these
various networks by using different routing algorithms. Pastry and
Tapestry use the algorithm of Plaxton~\etal~\cite{PlaxtonRR1997},
which is based on hypercube routing: the message is forwarded
deterministically to a neighbor whose identifier is one digit
closer
to the target identifier. CAN partitions a $d$-dimensional
coordinate space into {\em zones} that are owned by nodes
which store keys mapped to their zone.
Routing is done by greedily forwarding messages to the neighbor
closest to the target zone. Chord maps nodes and
resources to identities of $b$ bits placed around a {\em modulo
$2^b$ identifier circle} and each node maintains links to distances 
$2^0, 2^1 \ldots$ for greedy routing.  With $m$ machines in the
system, most of these
networks use $O(\log m)$ space and time for routing and
$O(\log m)$ time for node insertion (with the exception of Chord 
that takes $O(\log^2 m)$ time).
Because hashing destroys the ordering on keys, DHT systems do
not support queries that seek near matches to a key or keys within
a given range.

Some of these systems try to optimize performance by taking topology
into account. Pastry~\cite{RowstronD2001, CastroDHR2002} and
Tapestry~\cite{ZhaoKJ2001, ZhaoJK2002}
exploit geographical proximity by choosing the physically
closest node out of all the possible nodes with an appropriate
identifier prefix. In CAN~\cite{RatnasamyFHKS2001},
each node measures its round-trip
delay to a set of landmark nodes, and accordingly places itself
in the co-ordinate space to facilitate routing with respect to
geographic proximity. This last method is not fully self-organizing
and may cause imbalance in the distribution of nodes leading to
hot spots. Some methods to solve the nearest neighbor problem
for overlay networks can be seen in~\cite{HildrumKRZ2002}
and~\cite{KargerR2002}.

Some of these systems are partly resilient to random node
failures, but their performance may be badly impaired by adversarial
deletion of nodes. Fiat and Saia~\cite{FiatS2002} present a
network which is resilient to adversarial deletion of a
constant fraction of the nodes; some extensions of this result can
be seen in~\cite{SaiaFGKS2002, Datar2002}. However, they do not give 
efficient methods to dynamically maintain such a network.

TerraDir~\cite{SilaghiBK2002} is a recent system that provides locality
and maintains a hierarchical data structure using caching and replication.
There are as yet no provable guarantees on load balancing and
fault tolerance for this system.


\subsection{Our approach}
\label{section-approach}

The underlying structure of Chord, CAN, and similar DHTs resembles a
balanced tree in which balancing depends on the near-uniform
distribution of the output of the hash function.  So the costs of
constructing, maintaining, and searching these data structures is
closer to the $\Theta(\log n)$ costs of tree operations than the
$\Theta(1)$ costs of traditional hash tables.  But because keys are
hashed, DHTs can provide only hash table functionality.  Our approach
is to exploit the underlying tree structure to give tree
functionality, while applying a simple distributed balancing scheme to
preserve balance and distribute load.

We describe a new model for a peer-to-peer network based on 
a distributed data structure 
that we call a \buzz{skip graph}. 
This distributed data
structure has several benefits: Resource location
and dynamic node addition and deletion can be done in logarithmic 
time, and each node in a skip graph requires only logarithmic space 
to store information about its neighbors. More importantly, there
is no hashing of the resource keys, so related resources are present
near each other in a skip graph. This may be useful for certain 
applications such
as prefetching of web pages, enhanced browsing, and efficient
searching. Skip graphs also support \buzz{complex queries} such
as range queries, 
i.e., locating resources whose keys lie within a certain specified 
range\footnote{Skip graphs support complex queries along a single
dimension i.e., for one attribute of the resource, for example, its
name key.}. 
There has been some interest in supporting complex queries
in peer-to-peer-systems, and designing a system that supports range 
queries was posed as an open question~\cite{HarrenHHLSS2002}.  
Skip graphs are resilient to node failures:
a skip graph tolerates removal of a large fraction of its nodes chosen 
at random without becoming disconnected, and even the loss of an
$O(\frac{1}{\log n})$ fraction of the nodes chosen by an adversary 
still leaves most of the nodes in the largest surviving component.
Skip graphs can also be constructed without knowledge of
the total number of nodes in advance. In contrast, DHT systems such as 
Pastry and Chord require \emph{a priori}
knowledge about the size of the system or its keyspace.

The rest of the paper is organized as follows: we describe
skip graphs and algorithms for them in detail in 
Sections~\ref{section-skip-graphs} and~\ref{section-algorithms}. 
We describe the fault-tolerance properties and the repair mechanism 
for a skip graph in Sections~\ref{section-fault-tolerance} 
and~\ref{section-repair}. We discuss contention analysis and
some recent related work in Sections~\ref{section-congestion} 
and~\ref{section-related} respectively. Finally, we conclude in 
Section~\ref{section-conclusion}.


\subsection{Model}
\label{section-model}

We briefly describe the model for our algorithms. We assume 
a \buzz{message passing} environment in which all processes 
communicate with each other by sending messages over a 
communication channel. The system is \buzz{partially synchronous}, 
i.e., there is a fixed upper bound (time-out) on the transmission 
delay of a message. Processes can \buzz{crash},
i.e., halt prematurely, and crashes are permanent.
We assume that each message
takes at most unit time to be delivered and any internal processing
at a machine takes no time. 


\section{Skip graphs}
\label{section-skip-graphs}

A \buzz{skip list}, introduced by Pugh~\cite{Pugh1990}, 
is a randomized balanced tree data
structure organized as a tower of increasingly sparse linked lists.
Level $0$ of a skip list is a
linked list of all nodes in increasing order by key.
For each $i$ greater than $0$,
each node in level $i-1$ appears in level $i$
independently with some fixed probability $p$.  In a doubly-linked
skip list, each node stores a predecessor pointer and a 
successor pointer for each list in which it appears, for an average of
$\frac{2}{1-p}$ pointers per node.
The lists at the higher level act as ``express lanes'' that allow the
sequence of nodes to be traversed quickly.
Searching for a node with a particular key involves searching first
in the highest level, and repeatedly
dropping down a level whenever it becomes clear that the node is
not in the current level.  Considering the search path in reverse
shows that no more than $\frac{1}{1-p}$ nodes are searched on average
per level, giving an average search time of
$O\left(\log n \frac{1}{(1-p)\log \frac{1}{p}}\right)$ with $n$ nodes
at level $0$.
Skip lists have been extensively
studied~\cite{Pugh1990, PapadakisMP1990, Devroye1992, KirschenhoferP1994,
KirschenhoferMP1995},
and because they
require no global balancing operations are particularly useful in
parallel systems~\cite{GabarroMM1996, GabarroM1997}.

\begin{figure}[htp]
  \begin{center}
  \includegraphics[scale=0.45]{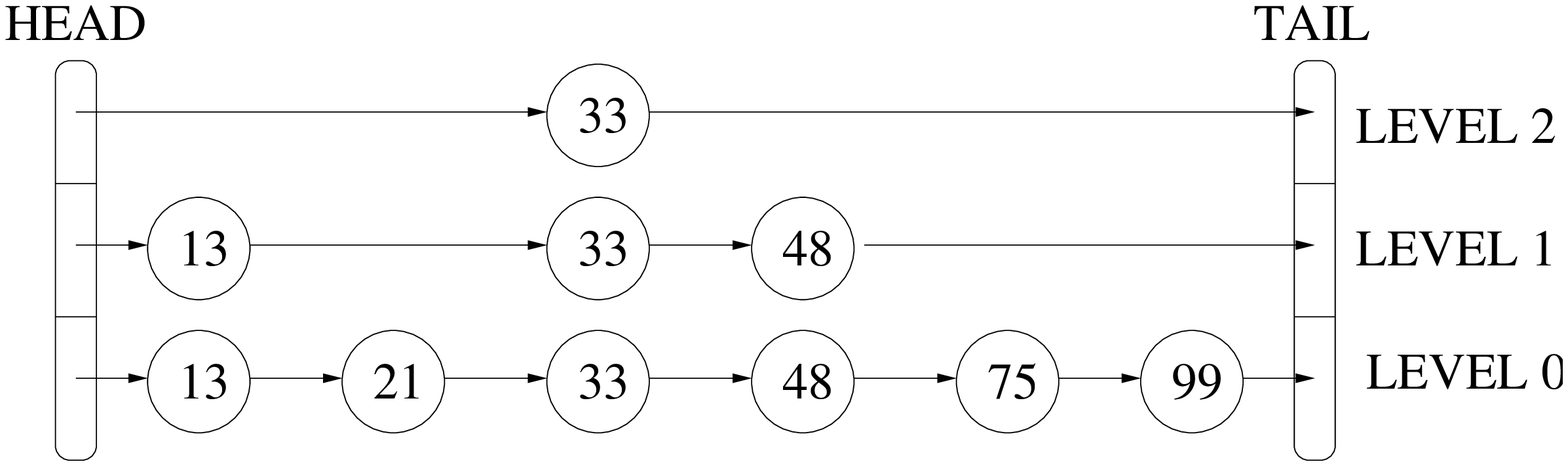}
  \end{center}
  \caption[A skip list.]
  {A skip list with $n=6$ nodes and $\lceil \log n \rceil = 3$ levels.}
  \label{fig-sl}
\end{figure}

We would like to use a data structure similar to a skip list to
support typical binary tree operations on a sequence whose nodes
are stored at separate locations in a highly distributed system subject to
unpredictable failures.  A skip list alone is not enough for our
purposes,
because it lacks redundancy and is thus vulnerable to both
failures and congestion. Since only a few nodes appear 
in the highest-level list, each such node acts as a single point of failure
whose removal partitions the list, and forms a hot spot that must
process a constant fraction of all search operations.
Skip lists also offer few guarantees
that individual nodes are not separated from the rest even with
occasional random failures.  Since each node is connected on average
to only $O(1)$ other nodes, even a constant probability of node
failures will isolate a large fraction of the surviving nodes.

Our solution is to define a generalization of a skip list that we call
a skip graph. 
As in a skip list, each of the $n$ nodes in a skip graph is
a member of multiple linked lists.  The level $0$ list
consists of all nodes in sequence.  Where a skip graph
is distinguished from a skip list is that there may be many
lists at level $i$, and every node participates in one of these lists,
until the nodes are splintered into singletons after $O(\log n)$
levels on average. A skip graph supports search, insert, and delete 
operations analogous to
the corresponding operations for skip lists; indeed, we show in
Lemma~\ref{lemma-restriction} that algorithms for skip lists can be
applied directly to skip graphs, as a skip graph is equivalent to a
collection of $n$ skip lists that happen to share some of their
lower levels.

Because there are many lists at each level, the chances that any
individual node participates in some search is small, eliminating both
single points of failure and hot spots.
Furthermore,
each node has $\Theta(\log
n)$ neighbors on average, and with high probability no node is
isolated.
In Section~\ref{section-fault-tolerance} we observe that skip
graphs are resilient to node failures and have an expansion ratio
of $\Omega(\frac{1}{\log n})$ with $n$ nodes in the graph.

In addition to providing fault tolerance, having an $\Omega(\log n)$
degree to support $O(\log n)$ search time appears to be necessary for
distributed data structures based on nodes in a one-dimensional space
linked by random connections satisfying certain uniformity
conditions~\cite{AspnesDS2002}.
While this lower bound requires some independence assumptions that are
not satisfied by skip graphs, there is enough similarity between skip
graphs and the class of models considered in the bound
that an $\Omega(\log n)$ average degree is not surprising.

We now give a formal definition of a skip graph.
Precisely which lists a node $x$ belongs to is controlled by a
\buzz{membership vector} $m(x)$.  We think of $m(x)$ as
an infinite random word over some fixed alphabet, although
in practice, only an
$O(\log n)$ length prefix of $m(x)$ needs to be generated on average.
The idea of the membership vector is that every linked list
in the skip graph is labeled by some finite word $w$, and
a node $x$ is in the list labeled by $w$ if and only if $w$ is a
prefix of $m(x)$.

\begin{figure}[htp]
  \begin{center}
  \includegraphics[scale=0.4]{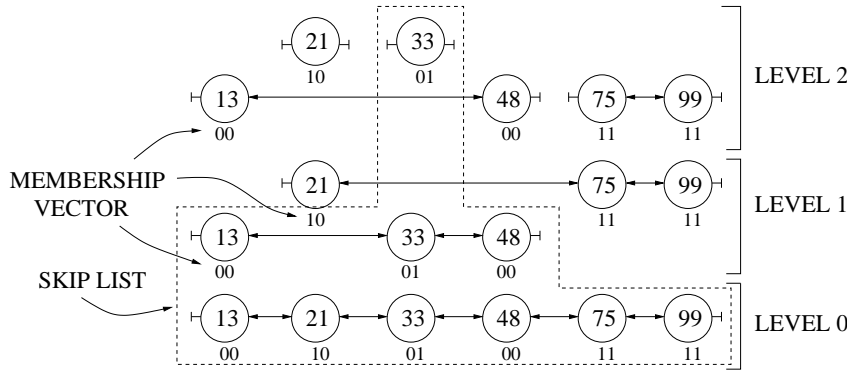}
  \end{center}
  \caption[A skip graph.]
  {A skip graph with $n=6$ nodes and $\lceil \log n \rceil = 3$ levels.}
  \label{fig-msl}
\end{figure}

To reason about this structure formally, we will need
some notation.
Let $\Sigma$ be a finite alphabet, let $\Sigma^*$ be the set of all
finite words consisting of characters in $\Sigma$, and let
$\Sigma^\omega$ consist of all infinite words.
We use subscripts to refer to individual characters of a word,
starting with subscript $0$; a word $w$ is equal to $w_0w_1w_2\ldots$.
Let $|w|$ be the length of $w$, with $|w| = \infty$ if $w \in \Sigma^\omega$.
If $|w| \ge i$,
write $w \restrictedto i$ for the prefix of $w$ of length $i$.
Write $\epsilon$ for the empty word. If $v$ and $w$ are both words,
write $v \prefix w$ if $v$ is a prefix of $w$,
i.e., if $w\restrictedto |v| = v$.
Write $w_i$ for the $i$-th character of the word $w$.
Write  $w_1\wedge w_2$ for the common prefix (possibly empty) of
the words $w_1$ and $w_2$.

Returning to skip graphs, the bottom level is always a 
doubly-linked list $S_\epsilon$ consisting of all the nodes in order
as shown in Figure~\ref{fig-msl}.
In general, for each $w$ in $\Sigma^*$, the doubly-linked
list $S_w$ contains all $x$ for which $w$ is a prefix of $m(x)$, in
increasing order.
We say that a particular list $S_w$ is part of level $i$ if $|w|=i$.
This gives an infinite family of doubly-linked lists; 
in an actual
implementation, only those $S_w$ with at least two nodes are
represented.
A skip graph is precisely a family $\{S_w\}$ of doubly-linked lists
generated in this fashion.
Note that because the membership vectors are random variables, each
$S_w$ is also a random variable.

We can also think of a skip graph as a random graph, where
there is an edge between $x$ and $y$ whenever $x$ and $y$ are
adjacent in some $S_w$.
Define $x$'s left and right neighbors at level $i$ as its immediate 
predecessor and successor, respectively, in $S_{m(x)\restrictedto i}$, 
or $\nil$ if no such nodes exist. We will write $xL_i$ for $x$'s left
neighbor at level $i$ and $xR_i$ for $x$'s right neighbor, and
in general will think of the $R_i$ as forming a family of 
associative composable operators to allow
writing expressions like $xR_iR_{i-1}^2$ etc. We write
$x.$maxLevel for the first level $\ell$ at which $x$ is in a
a singleton list, i.e., $x$ has at least one neighbor at 
level $\ell-1$.

An alternative view of a skip graph is a
\buzz{trie}~\cite{Briandais1959, Fredkin1960, Knuth1973}
of skip lists that share their lower levels. If we think of a skip
list formally as a sequence of random variables $S_0, S_1, S_2,
\ldots$, where the value of $S_i$ is the level $i$ list, then we
have:
\begin{lemma}
\label{lemma-restriction}
Let $\{S_w\}$ be a skip graph with alphabet $\Sigma$.
For any $z \in \Sigma^\omega$,
the sequence $S_0, S_1, S_2, \ldots$, where each
$S_i = S_{z \restrictedto i}$, is a skip list with parameter
$p=|\Sigma|^{-1}$.
\end{lemma}
\begin{proof}
By induction on $i$.  The list $S_0$ equals $S_\epsilon$, which is
just the base list of all nodes.
A node $x$ appears in $S_i$ if
$m(x) \restrictedto i = z \restrictedto i$;
conditioned on this event occurring, the probability that
$x$ also appears in $S_{i+1}$ is just the probability that
$m(x)_{i+i} = z_{i+1}$.
This event occurs with probability $p = |\Sigma|^{-1}$,
and it is easy to see that it is independent of the corresponding
event for any other $x'$ in $S_i$.
Thus each node in $S_i$ appears in $S_{i+1}$ with independent
probability $p$, and $S_0, S_1, \ldots$ form a skip list.
\end{proof}

For a node $x$ with membership vector $m(x)$, let
the skip list $S_{m(x)}$ be called the
\buzz{skip list restriction} of node $x$.


\subsection{Implementation}
\label{section-implementation}

In an actual implementation of a peer-to-peer system using a skip graph,
each node in a skip graph will be a resource. The resources are sorted
in increasing lexicographic order of their keys. Mapping
these keys to actual physical machines can be done in two ways:
In the first approach, we make every machine responsible for the resources 
that it hosts. Alternatively, we use a DHT approach where we hash node 
identifiers and resource keys to determine which nodes will be responsible 
for which keys. The first approach gives security and manageability 
whereas the second one gives good load balancing. 
For now, we treat nodes in the skip graph as representing
resources, and present our results without committing to 
how these resources are distributed across machines.
Each node in a skip graph stores the address and the key of its 
successor and predecessor at each of the $O(\log n)$ levels.
In addition, each node also needs $O(\log n)$ bits of space
for its membership vector. 

In both of the above approaches, with $n$ resources in the network,
each machine is responsible for maintaining $O(\log n)$ links
for {\em each} resource that it hosts, for a total of $O(n\log n)$
links in the entire network. This is a much higher storage 
requirement than the $O(m\log m)$ links for DHTs,
where $m$ is the number of machines in the system. 
Further, in our repair mechanism (described in Section~\ref{section-repair}), 
each machine will periodically check to see that its links are functional. 
This may result in a flood of messages given the high number of links per 
machine. It is an open question how to reduce the number of pointers 
in a skip graph and yet maintain the locality properties.


\section{Algorithms for a skip graph}
\label{section-algorithms}

In this section, we describe the search, insert and delete
operations for a skip graph. For simplicity, we refer to the key of a node
(e.g. $x.$key) with the same notation (e.g. $x$) as the node itself. 
It will be clear from the context whether we refer to a node or its key.
In the algorithms, we denote the pointer to $x$'s successor and predecessor
at level $\ell$ as \neigh{x}{R}{\ell} and \neigh{x}{L}{\ell} respectively.  
We define $xR_{\ell}$ formally to be the value of \neigh{x}{R}{\ell}, if 
\neigh{x}{R}{\ell} is a non-nil pointer to a non-faulty node, and $\nil$ 
otherwise. We define $xL_{\ell}$ similarly.  We summarize the 
variables stored at each node in Table~\ref{table-variables}.

\begin{table*}[htp]
\begin{center}
\begin{tabular}{|c|c|}
\hline
Variable&Type\\
\hline
key&Resource key\\
\hline
neighbor[$R$]&Array of successor pointers\\
\hline
neighbor[$L$]&Array of predecessor pointers\\
\hline
$m$&Membership vector\\
\hline
maxLevel&Integer\\
\hline
deleteFlag&Boolean\\
\hline
\end{tabular}
\end{center}
\caption{List of all the variables stored at each node.}
\label{table-variables}
\end{table*}

In this section, we only give the algorithms and analyze their
performance; we defer the proofs of the correctness of the 
algorithms to Section~\ref{section-correctness}.


\subsection{The search operation}
\label{section-search}

The search operation (Algorithm~\ref{alg-search})
is identical to the search in a skip list with only minor adaptations
to run in a distributed system. The search is started at the
topmost level of the node seeking a key and it proceeds along 
each level without overshooting the key, continuing at a lower
level if required, until it reaches level $0$. Either the address
of the node storing the search key, if it exists, or the address
of the node storing the largest key less than the search key is returned.


\begin{algorithm}[htp]
\caption{search for node $v$
\label{algorithm-search}}
\label{alg-search}
    \lnl{s1}
    upon receiving \msg{searchOp}{startNode, searchKey, level}:\;
      \lnl{s2}
      \uIf{($v$.key = searchKey)}
      {
         \lnl{s3}
         \KwSend \msg{foundOp}{$v$} to startNode\;
      }
      \lnl{s4}
      \uIf{($v$.key $<$ searchKey)}
      {
         \lnl{s5}
         \While{level $\geq 0$}
         {
           \lnl{s6}
           \uIf{((\neigh{v}{R}{\textrm{level}}.key $<$ searchKey)}
           {
             \lnl{s7}
             \KwSend \msg{searchOp}{startNode, searchKey, level} to \neigh{v}{R}{\textrm{level}}\;
             \lnl{s8}
             \KwBreak\;
           }
           \vspace*{5pt}
           \lnl{s9}
           \lElse
           {level$\leftarrow$level-1}\;
         }
      }
      \lnl{s10}
      \Else
      {
         \lnl{s11}
         \While{level $\geq 0$}
         {
           \lnl{s12}
           \uIf{((\neigh{v}{L}{\textrm{level}}).key $>$ searchKey)}
           {
              \lnl{s13}
              \KwSend \msg{searchOp}{startNode, searchKey, level} to \neigh{v}{L}{\textrm{level}}\;
              \lnl{s14}
              \KwBreak\;
           }
           \vspace*{5pt}
           \lnl{s15}
           \lElse{level$\leftarrow$level-1\;}
        }
      }
      \lnl{s16}
      \If{(level $<0$)}
      {
        \lnl{s17}
        \KwSend \msg{notFoundOp}{$v$} to startNode\;
      }
\end{algorithm}


\pagebreak
\begin{lemma}
\label{lemma-search-time} The search operation in a skip
graph $S$ with $n$ nodes takes expected $O(\log n)$ messages 
and $O(\log n)$ time.
\end{lemma}
\begin{proof}
Let $\Sigma$ be the alphabet for the membership vectors of the
nodes in the skip graph $S$, and
$z$ be the node at which the search starts. By
Lemma~\ref{lemma-restriction}, the sequence $S_{m(z)} = S_0,
S_1, S_2, \ldots$, where each $S_i =S_{z \restrictedto i}$, is a
skip list. A search that starts at $z$ in the skip graph will
follow the same path in $S$ as in $S_{m(z)}$. So we can directly 
apply the skip list search analysis given in~\cite{Pugh1990}, 
to analyze the search in $S$. With $n$ nodes, on an average there 
will be $O(\log n\frac{1}{\log (1/p)})$ levels, 
for $p=|\Sigma|^{-1}$. At most
$\frac{1}{1-p}$ nodes are searched on average at each level,
for a total of $O(\log n\frac{1}{(1-p)\log(1/p)})$ expected
messages and $O(\log n\frac{1}{(1-p)\log(1/p)})$ expected time. 
Thus, with fixed $p$, the search operation takes expected
$O(\log n)$ messages and $O(\log n)$ time.
\end{proof}


The network performance depends on the value of $p=|\Sigma|^{-1}$.
As $p$ increases, the search time decreases, but the
number of levels increase, so each node has to maintain 
neighbors at more levels. Thus we get a trade-off between the
search time and the storage requirements at each node. 

The performance shown in Lemma~\ref{lemma-search-time}
is comparable to the performance
of distributed hash tables, for example, Chord~\cite{StoicaMKKB2001}.
With $n$ resources in the system, a skip graph takes $O(\log n)$ time
for one search operation. In comparison, Chord takes $O(\log m)$ time,
where $m$ is the number of machines in the system. As long as
$n$ is polynomial in $m$, we get the same asymptotic performance from 
both DHTs and skip graphs for search operations.

Skip graphs can support \buzz{range queries} in which one is asked to
find a key $\geq x$, a key $\leq x$, the largest key $< x$, the least key
$> x$, some key in the interval $[x,y]$, all keys in $[x,y]$,
and so forth.
For most of these queries, the procedure is an obvious modification of
Algorithm~\ref{algorithm-search} and runs in $O(\log n)$ time with
$O(\log n)$ messages.
For finding all nodes in an interval, we can use a modified
Algorithm~\ref{algorithm-search} to find a single element of the
interval (which takes $O(\log n)$ time and $O(\log n)$ messages). With
$r$ nodes in the interval, we can then broadcast the query through all 
the nodes (which takes $O(\log r)$ time and $O(r \log n)$ messages).  
If the originator of
the query is capable of processing $r$ simultaneous responses, the
entire operation still takes $O(\log n)$ time.


\subsection{The insert operation}
\label{section-insert}

A new node $u$ knows some {\em introducing} node $v$ in the network 
that will help it to join the network. Node $u$ inserts itself 
in one linked list at each level till it finds itself in a singleton 
list at the topmost level. The insert operation consists of two stages:

\begin{enumerate}

  \item Node $u$ starts a search for itself from $v$ to find
        its neighbors at level $0$, and links to them.
  \item Node $u$ finds the closest nodes $s$ and $y$ at each level 
        $\ell \geq 0$, $s<u<y$, 
        such that $m(u)\restrictedto (\ell+1)=m(s)\restrictedto (\ell+1)
        = m(y) \restrictedto (\ell+1)$, if they exist, 
        and links to them at level $\ell+1$.
\end{enumerate}

\begin{figure}[htp]
  \begin{center}
  \includegraphics[scale=0.34]{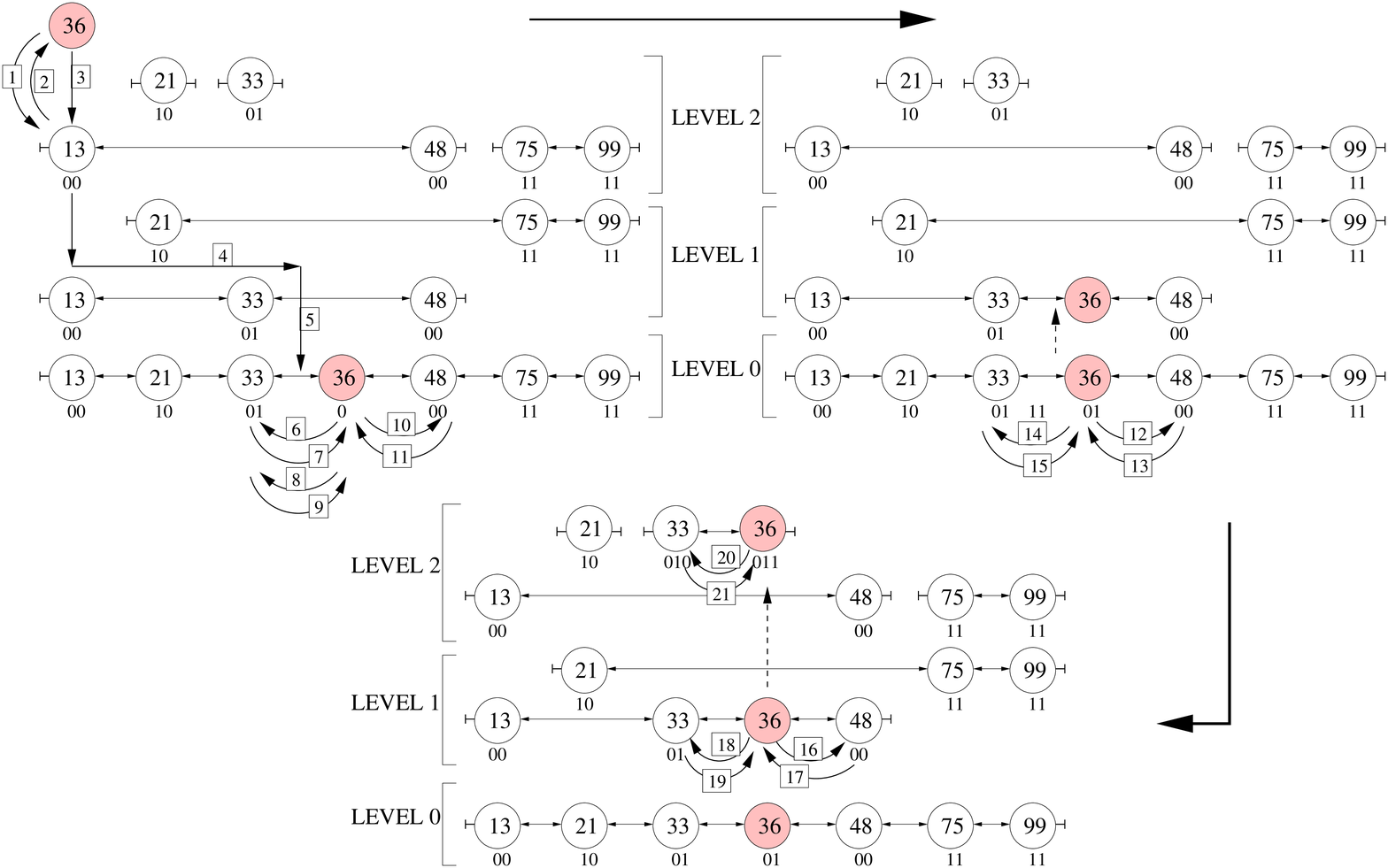}
  \end{center}
  \caption[Insert in a skip graph.]
  {Inserting node $36$ in a skip graph with $\Sigma=\{0,1\}$, starting from 
   node $13$. Messages are labeled by numbers
   in boxes in the order in which they are sent. Messages 1--3 implement
   node $36$ determining the maximum level of node $13$, and starting the search 
   operation to find its neighbor at level $0$. Messages 4--5 implement
   the search operation, and node $33$ informing node $36$ that it is node $36$'s closest
   neighbor at level $0$. Messages 6--11 implement node $36$ inserting itself
   between nodes $33$ and $48$ at level $0$. Messages 12--15 implement node $36$
   determining its neighbors at level $1$, and inserting itself between
   nodes $33$ and $48$ at level $1$. Messages 16--19 implement node $36$
   determining its neighbors at level $2$, and linking to node $33$ at level
   $2$. Messages 20--21 implement node $36$ determining its neighbors at
   level $3$, finding that no neighbors exist, and completing its insert 
   operation.}
  \label{fig-sg-insert}
\end{figure}

Because each existing node $v$ does not require $m(v)_{\ell+1}$
unless there exists another node $u$ such that 
$m(v)\restrictedto (\ell+1) = m(u)\restrictedto (\ell+1)$, 
it can delay determining its value
until a new node arrives asking for its value; thus at any given
time only a finite prefix of the membership vector of any node
needs to be generated. Detailed pseudocode for the insert operation
is given in Algorithm~\ref{alg-insert-new-node}.  
Figure~\ref{fig-sg-insert} shows a typical execution of an insert
operation in a small skip graph with $\Sigma=\{0,1\}$, where node 
$u=36$ is inserted starting from node $v=13$.  

Inserts can be trickier when we have to deal with concurrent node
joins. Before $u$ links to any neighbor, it verifies that its
join will not violate the order of the nodes. So if any new
nodes have joined the skip graph between $u$ and its
predetermined successor, $u$ will advance over the new nodes if
required before linking in the correct location.


\begin{algorithm}[htp]
\caption{insert for new node $u$}
\label{alg-insert-new-node}
    \lnl{in1}
    \uIf{(introducer = u)}
    {
        \lnl{in2}
        \neigh{u}{L}{0} $\leftarrow \nil$\;
        \lnl{in3}
        \neigh{u}{R}{0} $\leftarrow \nil$\;
        \lnl{in4}
	    $u.maxLevel \leftarrow 0$\;
    }
    \lnl{in5}
    \Else
    {
        \lnl{in6}
        \If{(introducer.key $< u$.key)}
        {
          \lnl{in7}
          side $\leftarrow R$\;
          \lnl{in8}
          otherSide $\leftarrow\; L$
        }
        \lnl{in9}
        \Else
        {
          \lnl{in10}
          side $\leftarrow L$\;
          \lnl{in11}
          otherSide $\leftarrow R$\;
        }
        \lnl{in12}
	\KwSend \msgnop{getMaxLevelOp} to introducer\;
        \lnl{in13}
        wait until receipt of \msg{retMaxLevelOp}{maxLevel}\;
        \lnl{in14}
        \KwSend \msg{searchOp}{$u$, $u$.key, maxLevel-1} to introducer\;
	\vspace*{5pt}
        \lnl{in15}
        wait until \op{foundOp} or \op{notFoundOp} is received\;
        \lnl{in16}
        upon receiving \msg{foundOp}{clone}:\;
        \lnl{in17}
          \ \ \ \ terminate insert\;
	\vspace*{5pt}
        \lnl{in18}
        upon receiving \msg{notFoundOp}{otherSideNeighbor}:\;
        \lnl{in19}
        \KwSend \msg{getNeighborOp}{side, $0$} to otherSideNeighbor\;
        \lnl{in20}
        wait until receipt of \msg{retNeighborOp}{sideNeighbor, $0$}:\;
        \lnl{in21}
        \KwSend \msg{getLinkOp}{$u$, side, $0$} to otherSideNeighbor\;
        \lnl{in22}
        wait until receipt of \msg{setLinkOp}{newNeighbor, $0$}:\;
        \lnl{in23}
        \neigh{u}{\textrm{otherSide}}{0} $\leftarrow$ newNeighbor\;
        \lnl{in24}
        \KwSend \msg{getLinkOp}{$u$, otherSide, $0$} to sideNeighbor\;
        \lnl{in25}
        wait until receipt of \msg{setLinkOp}{newNeighbor, $0$}:\;
        \lnl{in26}
        \neigh{u}{\textrm{side}}{0} $\leftarrow$ newNeighbor\;
        \vspace*{5pt}
        \lnl{in27}
        $\ell \leftarrow 0$\;
        \lnl{in28}
        \While{true}
        {
            \lnl{in29}
            $m(u)_{\ell} \leftarrow$ uniformly chosen random element of $\Sigma$\;
            \lnl{in30}
            $\ell \leftarrow \ell+1$\;
            \lnl{in31}
            \uIf{(\neigh{u}{R}{\ell-1} $\neq \nil$)}
            {
               \lnl{in32}
               \KwSend \msg{buddyOp}{$u$, $\ell-1$, $m(u)_{\ell-1}$, $L$}
                     to \neigh{u}{R}{\ell-1}\;
               \lnl{in33}
               wait until receipt of \msg{setLinkOp}{neighbor, $\ell$}:\;
               \lnl{in34}
               \neigh{u}{R}{\ell} $\leftarrow$ neighbor\;
            }
            \lnl{in35}
            \lElse{\neigh{u}{R}{\ell} $=\nil$\;}
            \lnl{in36}
            \If{(\neigh{u}{L}{\ell-1} $\neq \nil$)}
            {
               \lnl{in37}
               \KwSend \msg{buddyOp}{$u$, $\ell-1$, $m(u)_{\ell-1}$, $R$}
                     to \neigh{u}{L}{\ell-1}\;
               \lnl{in38}
               wait until receipt of \msg{setLinkOp}{neighbor, $\ell$}:\;
               \lnl{in39}
               \neigh{u}{L}{\ell} $\leftarrow$ neighbor\;
            }
            \lnl{in40}
            \lElse{\neigh{u}{L}{\ell} $=\nil$\;}
            \lnl{in41}
            \If{((\neigh{u}{R}{\ell} $=\nil$) {\bf and}
                (\neigh{u}{L}{\ell} $=\nil$))}
            {
               \lnl{in42}
	       \KwBreak\;
	    }
        }
        \lnl{in43}
	$u.\textrm{maxLevel}\leftarrow \ell$\;
    }
\end{algorithm}


\begin{algorithm}[htp]
\caption{Node $v$'s message handler for messages received during the insert
         of new node $u$.}
\label{alg-insert-old-node}
    \lnl{io1}
    upon receiving \msg{getLinkOp}{$u$, side, $\ell$}:\;
    \lnl{io2}
    change\_neighbor($u$, side, $\ell$)\;

    \vspace*{5pt}
    \lnl{io3}
    upon receiving \msg{buddyOp}{$u$, $\ell$, val, side}:\;
       \lnl{io4}
       \lIf{(side $= L$)}
       {
         otherSide $\leftarrow R$\;
       }
       \lnl{io5}
       \lElse
       {
         otherSide $\leftarrow L$\;
       }
       \lnl{io6}
       \If{($m(v)_{\ell} = \nil$)}
       {
          \lnl{io7}
          $m(v)_{\ell}\leftarrow$ uniformly chosen random element of $\Sigma$\;
          \lnl{io8}
          \neigh{v}{L}{\ell} $\leftarrow \nil$\;
          \lnl{io9}
          \neigh{v}{R}{\ell} $\leftarrow \nil$\;
       }
       \lnl{io10}
       \uIf{($m(v)_{\ell} = val$)}
       {
         \lnl{io11}
         change\_neighbor($u$, side, $\ell+1$)\;
       }
       \lnl{io12}
       \Else
        {
          \lnl{io13}
          \uIf{(\neigh{v}{\textrm{otherSide}}{\ell} $\neq \nil$)}
          {
             \lnl{io14}
             \KwSend \msg{buddyOp}{$u$, val, $\ell$, side} 
                     to \neigh{v}{\textrm{otherSide}}{\ell}\;
          }
          \lnl{io15}
          \Else
          {
            \lnl{io16}
            \KwSend \msg{setLinkOp}{$\nil$, $\ell$} to $u$\;
          }
        }
\end{algorithm}


\begin{algorithm}[htp]
\caption{change\_neighbor($u$, side, $\ell$) for node $v$}
\label{alg-change-neigh}
       \lnl{cn1}
       \lIf{(side $=R$)}
       {
          cmp $\leftarrow\  <$\;
       }
       \lnl{cn2}
       \lElse
       {
          cmp $\leftarrow\  >$\;
       }
       \lnl{cn3}
       \uIf{((\neigh{v}{\textrm{side}}{\ell}).key cmp $u$.key)}
          {
            \lnl{cn4}
            \KwSend \msg{getLinkOp}{$u$, side, $\ell$} to \neigh{v}{\textrm{side}}{\ell}\;
          }
       \lnl{cn5}
       \Else
       {
         \lnl{cn6}
         {\KwSend \msg{setLinkOp}{$v$, $\ell$} to $u$}\;
       }
       \lnl{cn7}
       \neigh{v}{\textrm{side}}{\ell} $\leftarrow u$\;
\end{algorithm}


\begin{algorithm}[htp]
\caption{Additional messages for node $v$}
\label{alg-update}
  \lnl{u1}
  upon receiving \msg{updateOp}{side, newNeighbor, $\ell$}:\;
  \lnl{u2}
  \neigh{v}{\textrm{side}}{\ell} $\leftarrow$ newNeighbor\;
  \vspace*{5pt}
  \lnl{u3}
  upon receiving \msgnop{getMaxLevelOp} from $u$:\;
  \lnl{u4}
  \KwSend \msg{retMaxLevelOp}{$v.maxLevel$} to $u$\;
  \vspace*{5pt}
  \lnl{u5}
  upon receiving \msg{getNeighborOp}{side, $\ell$} from $u$:\;
  \lnl{u6}
  \KwSend \msg{retNeighborOp}{$v\textrm{side}_{\ell}$} to $u$\;
\end{algorithm}


\begin{lemma}
\label{lemma-insert-time} The insert operation in a skip
graph $S$ with $n$ nodes takes expected $O(\log n)$ 
messages and $O(\log n)$ time.
\end{lemma}
\begin{proof}
Let $\Sigma$ be the alphabet for the membership vectors of the
nodes in the skip graph $S$. With $n$ nodes, there will be average 
of $O(\log n\frac{1}{\log (1/p)})$ levels in the skip graph, 
$p=|\Sigma|^{-1}$. To link at level $0$, a new node $u$ performs 
one search operation. From Lemma~\ref{lemma-search-time}, this 
takes $O(\log n\frac{1}{(1-p)\log(1/p)})$ expected messages and 
$O(\log n\frac{1}{(1-p)\log(1/p)})$ expected time.
At each level $\ell$, $\ell\geq0$, $u$ communicates
with an average of ${2}/{p}$ nodes,
before it finds at most two nodes $s$ and $y$, 
with $m(s)_{\ell}=m(u)_{\ell}=m(y)_{\ell}$, $s<u<y$, and connects 
to them at level $\ell+1$. The expected number of messages and time
for the insert operation at all levels is 
$O\left(\frac{\log n}{\log (1/p)}\left(\frac{1}{1-p} +\frac{2}{p}
\right)\right)$. Thus with fixed $p$, the insert operation takes 
expected $O(\log n)$ messages and $O(\log n)$ time.
\end{proof}

With $m$ machines and $n$ resources in the system,
most DHTs such as CAN, Pastry and Tapestry take $O(\log m)$ time
for insertion; an exception is Chord which takes $O(\log^2 m)$ time.
An $O(\log m)$ time bound improves on the $O(\log n)$ bound for
skip graphs when $m$ is much smaller than $n$. However, the cost
of this improvement  is losing support for complex queries and 
spatial locality, and the improvement itself is only a constant 
factor unless some machines store a superpolynomial number of 
resources.


\subsection{The delete operation}
\label{section-delete}

The delete operation is very simple. When node $u$ wants to leave
the network, it informs its predecessor node at each level to 
update its successor pointer to point to $u$'s successor. 
It starts at the topmost level and works its way
down to level $0$. Node $u$ also informs its successor node at each
level to update its predecessor pointer to point to $n$'s predecessor.
If $u$'s successor or predecessor are being deleted as well,
they pass the message on to their neighbors so that the nodes
are correctly linked up. A node does not delete itself from
the graph as long as it is waiting for some message as a part 
of the delete operation of another node.


\begin{algorithm}[htp]
\caption{delete for existing node $u$}
\label{alg-delete-new}
  
  \lnl{dn1}
  $u.$deleteFlag = true\;
  \lnl{dn2}
  \For{$\ell\leftarrow u$.max\_levels \KwDownTo $0$}
  {
    \lnl{dn3}
    \If{\neigh{u}{R}{\ell} $\neq \nil$}
    {
        \lnl{dn4}
        \KwSend \msg{deleteOp}{$\ell$, sender} to \neigh{u}{R}{\ell}\;
        \lnl{dn5}
        wait until receipt of \msg{confirmDeleteOp}{$\ell$} or \msg{noNeighborOp}{$\ell$}:\;
        \lnl{dn6}
        upon receiving \msg{noNeighborOp}{$\ell$}:\;
        \lnl{dn7}
        \If{\neigh{u}{L}{\ell} $\neq \nil$}
        {
            \lnl{dn8}
            \KwSend \msg{setNeighborNilOp}{$\ell$, sender} to \neigh{u}{L}{\ell}\;
            \lnl{dn9}
            wait until receipt of \msg{confirmDeleteOp}{$\ell$}\;
        }
    }
  }
\end{algorithm}


\begin{algorithm}[htp]
\caption{Node $v$'s message handler for messages received during the delete operation.}
\label{alg-delete-other}
  \lnl{do1}
  upon receiving \msg{deleteOp}{$\ell$, sender}:\;
  \lnl{do2}
  \If{($v.$deleteFlag = true)}
  {
      \lnl{do3}
      \If{(\neigh{v}{R}{\ell} $\neq \nil$)}
      {
          \lnl{do4}
          \KwSend \msg{deleteOp}{$\ell$, sender} to \neigh{v}{R}{\ell}\;
      }
      \lnl{do5}
      \Else
      {
          \lnl{do6}
          \KwSend \msg{noNeighborOp}{$\ell$} to sender\;
      }
  }
  \lnl{do7}
  \Else
  {
      \lnl{do8}
      \KwSend \msg{findNeighborOp}{$\ell$, sender} to \neigh{v}{L}{\ell}\;
      \lnl{do9}
      wait until receipt of \msg{foundNeighborOp}{$x$, $\ell$}:\;
      \lnl{do10}
      \neigh{v}{L}{\ell} $\leftarrow x$\;
      \lnl{do11}
      \KwSend \msg{confirmDeleteOp}{$\ell$} to sender\;
  }

  \vspace*{5pt}
  \lnl{do12}
  upon receiving \msg{findNeighborOp}{$\ell$, sender}:\;
  \lnl{do13}
  \If{($v.$deleteFlag = true)}
  {
     \lnl{do14}
     \If{(\neigh{v}{L}{\ell} $\neq \nil$)}
     {
         \lnl{do15}
         \KwSend \msg{findNeighborOp}{$\ell$, sender} to \neigh{v}{L}{\ell}\;
     }
     \lnl{do16}
     \Else
     {
         \lnl{do17}
         \KwSend \msg{foundNeighborOp}{$\nil$, $\ell$} to sender\;
     }
  }
  \lnl{do18}
  \Else
  {
      \lnl{do19}
      \KwSend \msg{foundNeighborOp}{$v$, $\ell$} to sender\;
      \lnl{do20}
      \neigh{v}{R}{\ell} $\leftarrow$ sender\;
  }

  \vspace*{5pt}
  \lnl{do21}
  upon receiving \msg{setNeighborNilOp}{$\ell$, sender}:\;
  \lnl{do22}
  \If{($v.$deleteFlag = true)}
  {
      \lnl{do23}
      \If{(\neigh{v}{L}{\ell} $\neq \nil$)}
      {
          \lnl{do24}
          \KwSend \msg{setNeighborNilOp}{$\ell$, sender} to \neigh{v}{L}{\ell}\;
      }
      \lnl{do25}
      \Else
      {
          \lnl{do26}
          \KwSend \msg{confirmDeleteOp}{$\ell$} to sender\;
      }
  }
  \lnl{do27}
  \Else
  {
      \lnl{do28}
      \KwSend \msg{confirmDeleteOp}{$\ell$} to sender\;
      \lnl{do29}
      \neigh{v}{R}{\ell} $\leftarrow \nil$\;
  }
\end{algorithm}


\begin{lemma}
\label{lemma-delete-time} The delete operation in a skip
graph $S$ with $n$ nodes takes expected $O(\log n)$ messages and
$O(1)$ time.
\end{lemma}
\begin{proof}
Let $\Sigma$ be the alphabet for the membership vectors of the
nodes in the skip graph $S$. With $n$ nodes, there will be average 
of $O(\log n\frac{1}{\log (1/p)})$ levels in the skip graph, 
$p=|\Sigma|^{-1}$. At each level $\ell$, $\ell\geq0$, 
the node to be deleted communicates with at most two other nodes.
It takes an average of $O(\log n\frac{1}{\log(1/p)})$ total messages and 
$O(1)$ time as the messages at all the levels can be sent in parallel.
Thus with fixed $p$, a delete operation takes $O(\log n)$ messages and
$O(1)$ time.
\end{proof}

During the delete operation of node $u$, if $u$'s successor
or predecessor at some level are also being deleted, then the 
number of message at that level is proportional to the number 
of consecutive nodes being deleted.


\subsection{Correctness of algorithms}
\label{section-correctness}

In this section, we prove the correctness of the insert and
delete algorithms given in Section~\ref{section-algorithms}.
The definition of a skip graph in Section~\ref{section-skip-graphs}
involves global properties of the data structure (such as 
$S_{w1}$ being a subset of $S_w$) that are difficult to work
with in the correctness proofs. So we start by defining a set
of {\em local} constraints which characterize a skip graph.
We first prove that a data structure is a skip 
graph if and if only if all these constraints are satisfied,
and then we prove that these constraints are not violated
after an insert and delete operation, thus maintaining the
skip graph properties. Further, these constraints will
be used for our repair mechanism as we can monitor the
state of the graph by checking these constraints locally
at each node, and detecting and repairing node failures.

As explained in Section~\ref{section-algorithms}, we use 
$\nil$ both to refer to the null pointers at the ends of 
the doubly-linked lists of the skip graph, and to refer to 
pointers to failed nodes. 

\pagebreak
Let $x$ be any node in the skip graph; then for all levels 
$\ell\geq 0$:
\begin{enumerate}
  \item If $xR_{\ell} \neq \nil$, $xR_{\ell} > x$.
  \label{c-right-bigger}

  \item If $xL_{\ell} \neq \nil$, $xL_{\ell} < x$.
  \label{c-left-smaller}

  \item If $xR_{\ell} \neq \nil$, $xR_{\ell}L_{\ell} = x$.
  \label{c-right-left}

  \item If $xL_{\ell} \neq \nil$, $xL_{\ell}R_{\ell} = x$.
  \label{c-left-right}

  \item If $m(x)\restrictedto (\ell+1) = m(xR_{\ell}^{k}) \restrictedto (\ell+1)$
        and $\nexists j, j<k, m(x)\restrictedto (\ell+1) = m(xR_{\ell}^j) 
        \restrictedto (\ell+1)$, then $xR_{\ell+1} = xR_{\ell}^{k}$. Else,
        $xR_{\ell+1}=\nil$.
  \label{c-right-successor}

  \item If $m(x)\restrictedto (\ell+1) = m(xL_{\ell}^{k}) \restrictedto (\ell+1)$
        and $\nexists j, j<k, m(x)\restrictedto (\ell+1) = m(xL_{\ell}^j) 
        \restrictedto (\ell+1)$, then $xL_{\ell+1} = xL_{\ell}^{k}$. Else,
        $xL_{\ell+1}=\nil$.
  \label{c-left-successor}
\end{enumerate}

As per the definition of a skip graph given in 
Section~\ref{section-skip-graphs}, all the elements in a doubly 
linked list $S_w$ (which contains all $x$ for which $w$ is a prefix 
of $m(x)$ of length $\ell$) are in increasing order. 
Constraints~\ref{c-right-bigger}
through~\ref{c-left-right} imply that all non-edge nodes satisfy the 
increasing order in the linked lists. Constraints~\ref{c-right-bigger} 
and~\ref{c-left-smaller} ensure that the order is locally true at every 
node, whereas Constraints~\ref{c-right-left} and~\ref{c-left-right} 
ensure that the entire list is doubly linked correctly. The
constraints that the increasing order is satisfied locally at each
node and that the list is doubly-linked correctly, put together ensure 
that no element is skipped over and that the entire list is sorted.

Constraints~\ref{c-right-successor} and~\ref{c-left-successor} 
denote how the lists at
different levels are related to each other. The successor
(predecessor) of node $x$ at level $\ell+1$ is always the first 
node to its right (left) at level $\ell$ whose membership vector 
matches the membership vector of $x$ in one additional position.
Node $x$ is connected at level $\ell+1$, on the right side 
to a node $z$ such that $x, z \in S_w$, and $z$ is the nearest 
node greater than $x$ in $S_w$ with $m(x)_{\ell} = m(z)_{\ell}$. 
Similarly, $x$ is connected at level $\ell+1$, on the left side 
to a node $u$ such that $u, x \in S_w$, and $u$ is the nearest 
node less than $x$ in $S_w$ with $m(u)_{\ell} = m(x)_{\ell}$. 

Define a {\em defective} skip graph as a data structure that 
that contains skip graph elements but does not satisfy the
definition of a skip graph; for example, it may contain 
out-of-order elements, missing links, or worse.


\begin{theorem}
\label{theorem-iff-skip-graph}
Every connected component of the data structure is a skip graph
if and only if Constraints $1-6$ are satisfied.
\end{theorem}
\begin{proof}
We start with the reverse direction: if the constraints are not
satisfied, then some connected component of the data structure is
not a skip graph. As 
Constraints~\ref{c-left-smaller},~\ref{c-left-right} 
and~\ref{c-left-successor} are mirror images of
Constraints~\ref{c-right-bigger},~\ref{c-right-left}
and~\ref{c-right-successor} respectively, we will only consider
violations of Constraints~\ref{c-right-bigger},~\ref{c-right-left}
and~\ref{c-right-successor}.

\begin{figure}[htp]
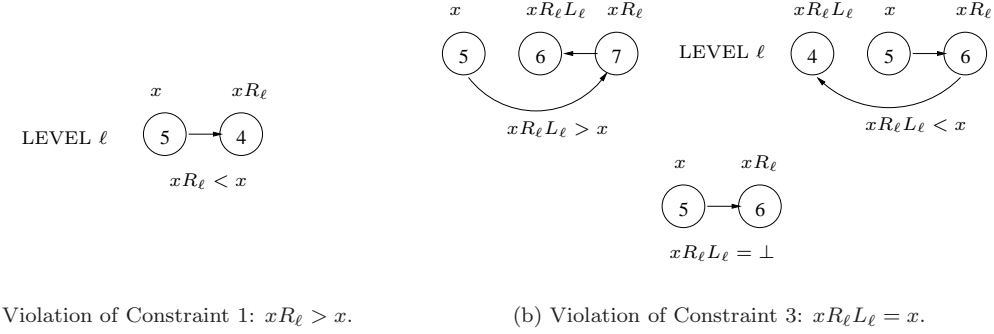

  \begin{center}
    \mbox{
    \subfigure[Violation of Constraint~\ref{c-right-bigger}: $xR_{\ell} > x$.] 
              {\input{violate-1.pstex_t}}\quad\quad
    \subfigure[Violation of Constraint~\ref{c-right-left}: $xR_{\ell}L_{\ell}=x$.] 
              {\input{violate-3.pstex_t}}
     }
    \caption{Violation of Constraints~\ref{c-right-bigger} and~\ref{c-right-left}.}
    \label{figure-violate}
  \end{center}
\end{figure}

Figure~\ref{figure-violate} shows how
Constraints~\ref{c-right-bigger} and~\ref{c-right-left} can
be violated. Each violation leads to either an {\em unsorted} or
{\em inconsistently} linked list at level $\ell$, so the data structure 
is not a skip graph. There are two ways in which 
Constraint~\ref{c-right-successor} can be violated.
\begin{enumerate}
  \item For some $x$ and $\ell$, $xR_{\ell+1}=xR_{\ell}^{k}$ but $\exists j, j<k, 
        m(x)\restrictedto (\ell+1) = m(xR_{\ell}^j) \restrictedto (\ell+1)$.\\
        Let $y=xR_{\ell}^j$ and $z=xR_{\ell}^{k}$. 
        As the linked list is sorted at level $\ell$,
        $j<k \Rightarrow y<z$, and since $y = xR_{\ell}^j, x<y$. Let 
        $S_w = \{x | m(x)\restrictedto (\ell+1) = w\}$.
        Then in a skip graph, $x, y$ and $z \in S_w$.
        Since $y \neq xR_{\ell+1}$, either $y \notin S_w$ or the linked list at 
        level $\ell+1$ is not sorted as $x<y<z$. In both cases, the resulting 
        data structure is a defective skip graph.
  \item For some $x$ and $\ell$, $xR_{\ell+1} \neq xR_{\ell}^{k}$.\\ 
        As $m(x) \restrictedto (\ell+1) = m(xR_{\ell+1}) \restrictedto (\ell+1)$,
        $m(x) \restrictedto \ell = m(xR_{\ell+1}) \restrictedto \ell$. It follows that 
        both $x$ and $xR_{\ell+1}$ are in $S_{m(x)\restrictedto \ell}$. But then if
        $xR_{\ell+1} \neq xR_{\ell}^k$ for any $k$, some edge in 
        $S_{m(x)\restrictedto {\ell}}$ is missing, and the data structure is a
        defective skip graph.
\end{enumerate}

Now we prove that every connected component of a data structure is a skip graph 
if all the constraints are satisfied. We first prove that we have sorted, 
doubly-linked lists at all levels using Constraints~\ref{c-right-bigger} 
through~\ref{c-left-right}, and then prove that each list contains the correct 
elements as per their membership vectors using Constraints~\ref{c-right-successor} 
and~\ref{c-left-successor}.

Let $x$ be an arbitrary element of the data structure. Let $S_{x,\ell}$
be the maximal sequence of the form $xL_{\ell}^j,\ldots,x,\ldots,xR_{\ell}^k$,
where $j, k\geq 0$, 
such that no element of the sequence is $\nil$. We show that this
sequence is sorted and doubly-linked using induction.
According to Constraint~\ref{c-right-bigger}, $xR_{\ell}>x$, and
according to Constraint~\ref{c-right-left}, $xR_{\ell}L_{\ell}=x$.
Thus the sequence $x,xR_{\ell}$ is sorted and doubly-linked.
Similarly, according to Constraint~\ref{c-left-smaller}, $xL_{\ell}<x$,
and according to Constraint~\ref{c-left-right}, $xL_{\ell}R_{\ell}=x$.
Thus the sequence $xL_{\ell},x$ is sorted and doubly-linked.
Let the sequence $xL_{\ell}^{j-1}, \ldots, x, \ldots, xR_{\ell}^{k-1}$, 
be a sorted, doubly-linked list. According to Constraints~\ref{c-right-bigger} 
and~\ref{c-right-left}, $xR_{\ell}^{k}>xR_{\ell}^{k-1}$, and 
$xR_{\ell}^{k-1}R_{\ell}L_{\ell}=xR_{\ell}^{k-1}$. Similarly, according 
to constraints~\ref{c-left-smaller} and~\ref{c-left-right},  
$xL_{\ell}^{j}<xL_{\ell}^{j-1}$, and 
$xL_{\ell}^{j-1}R_{\ell}L_{\ell}=xL_{\ell}^{j-1}$. Thus the maximal
sequence $S_{x,\ell} = xL_{\ell}^j,\ldots,x,\ldots,xR_{\ell}^k$ 
is sorted and doubly-linked. 

We now show that if two nodes are connected at level $\ell$, 
they are also connected at level $0$.
Suppose that Constraints~\ref{c-right-successor} and~\ref{c-left-successor}
hold. Then, we prove that for each level $\ell\geq0$, $xR_{\ell}=xR_0^j$
and $xL_{\ell}=xL_0^j$. Clearly this is true for $\ell=0$ and $j=1$.
Suppose that it is true for level $\ell-1$.
Let $x=y_0$ and let each $xR_{\ell-1}^i=y_i$, $1\leq i\leq k$. 
For each $i$, $y_i=y_{i-1}R_{\ell-1}=y_{i-1}R_0^{j_i}$. So
$y_1=y_0R_{\ell-1}=y_0R_0^{j_0}$, 
$y_2=y_1R_{\ell-1}=y_1R_0^{j_1}=y_0R_0^{j_0+j_1}$, and so on.
Thus, $y_k=y_0R_0^{j_0+j_1+j_2+\ldots+j_k}=y_0R_0^{j}$,
where $j=j_0+j_1+\ldots+j_k$. But $y_k=xR_{\ell-1}^k$ 
and $y_0=x$. So $xR_{\ell-1}=xR_0^j$. According to Constraint~\ref{c-right-successor}, 
$xR_{\ell}=xR_{\ell-1}^k$. Thus we get $xR_{\ell}=xR_0^j$. A
similar proof will show that $xL_{\ell}=x_0L_0^j$.

We use the proof above to show that any two connected nodes
are connected in the same list at level $0$.
Consider a path $x E_1 E_2 \ldots E_k y$ where each $E_i$ is
either $L_{\ell_i}$ or $R_{\ell_j}$. As proved above, there
exists a path $x E_1'^{j_1} E_2'^{j_2} \ldots E_k'^{j_k} y$
where each $E_i'$ is either $L_0$ or $R_0$. Thus it follows
that $x$ and $y$ are in the same list at level $0$. Also
$S_{x,0}=S_{y,0}$ if $x$ and $y$ are in the same connected
component. So we get a single list $S_{\epsilon}$ at level $0$,
which consists of all the elements in the same connected 
component of the data structures.

As proved above, $S_{\epsilon}$ is also sorted and doubly-linked.  
With the single list $S_{\epsilon}$ at level $0$,
according to Constraints~\ref{c-right-successor}
and~\ref{c-left-successor}, each node 
$x \in S_{\epsilon}$, is linked to its right and left at level $1$ to the 
nearest nodes $z$ and $u$ respectively (if they exist), 
such that $m(x)\restrictedto 1= m(z)\restrictedto 1=m(u)\restrictedto 1$,
and $u<x<z$. Thus, we get $|\Sigma|$ linked lists at level $1$, 
$S_a=\{y|m(y)_0=a\}$, one for each $a\in \Sigma$. In general, at level $\ell$, 
we can get up to $|\Sigma|^{\ell}$ lists, one for each $w\in\Sigma^{\ell}$. 
Each list contains all the nodes which have the matching membership
vector prefix. As proved above, each of these lists is also sorted and
doubly-linked. Thus, if the data
structure satisfies all the constraints, it is a skip graph.
\end{proof}


\begin{lemma}
\label{lemma-insert-preserve}
Inserting a new node $u$ in a skip graph $S$ using 
Algorithm~\ref{alg-insert-new-node} gives a skip graph.
\end{lemma}

\begin{proof}
Inserting a new node $u$ in $S$ consists of two stages:
inserting $u$ in level $0$ using a search operation,
and inserting $u$ in levels $\ell>0$ using the neighbors of
$u$ at level $\ell-1$.  We consider the case where the
introducing node's key is less than $u$'s key; the other
case is similar so we omit those details here. Also,
we only prove that 
Constraints~\ref{c-right-bigger},~\ref{c-right-left}
and~\ref{c-right-successor} are satisfied as 
Constraints~\ref{c-left-smaller},~\ref{c-left-right}
and~\ref{c-left-successor} are mirror images of
Constraints~\ref{c-right-bigger},~\ref{c-right-left}
and~\ref{c-right-successor} respectively.

The search operation started by $u$ (line~\ref{in14} 
of Algorithm~\ref{alg-insert-new-node}) returns the largest
node $s$ less than $u$ (line~\ref{s17} of 
Algorithm~\ref{alg-search}), and $u$ sends a \op{getLinkOp}
message to $s$ (line~\ref{in21} of 
Algorithm~\ref{alg-insert-new-node}). One of the following
two cases occur: Either $s$ sets $sR'_0=u>s$ (line~\ref{cn7} of
Algorithm~\ref{alg-change-neigh}), maintaining
Constraint~\ref{c-right-bigger}. Or, if additional
nodes are inserted between $u$ and $s$, and
$sR_0<u$ (line~\ref{cn3} of Algorithm~\ref{alg-change-neigh}),
then $s$ passes the \op{getLinkOp} message to $sR_0$.
As the \op{getLinkOp} message is only passed to nodes 
whose key is less than that of $u$, eventually it
reaches some node $t$ where the message terminates 
and $t'R_0=u>t$, maintaining Constraint~\ref{c-right-bigger}.
Also as $u$ sets $uL_0'=s<u$ or $uL_0'=t<u'$ 
(line~\ref{in23} of Algorithm~\ref{alg-insert-new-node}),
either $sR_0'L_0'=s$ or $tR_0'L_0'=t$ satisfying 
Constraint~\ref{c-right-left}.
In the absence of a suitable $s$, no pointers are changed and
Constraints~\ref{c-right-bigger} and~\ref{c-right-left} are 
satisfied as the pointer values remain unchanged from before the
insert.

Node $u$ also determines the initial right neighbor of $s$, say $z$
(lines~\ref{in19} and~\ref{in20} of Algorithm~\ref{alg-insert-new-node})
and sends a \op{getLinkOp} message to $z$ (line~\ref{in24} of
Algorithm~\ref{alg-insert-new-node}). Similar to the
earlier \op{getLinkOp} message, either $z$ sets $z'L_0=u<z$ (line~\ref{cn7}
of Algorithm~\ref{alg-change-neigh}), or passes the message
on to $zL_0$ if it is greater than $u$ (line~\ref{cn3} of
Algorithm~\ref{alg-change-neigh}). As the \op{getLinkOp} message is 
only passed to nodes whose key is greater than that of $u$, eventually 
it reaches some node $y$ where the message terminates and $y'L_0=u<y$. 
Node $u$ sets $u'R_0=z>u$ or $u'R_0=y>u$ (line~\ref{in26} of 
Algorithm~\ref{alg-insert-new-node}), maintaining 
Constraint~\ref{c-right-bigger}. As $z$ sets $zL_0'=u<z$ or
$y$ sets $yL_0'=u<y$ (line~\ref{cn7} of Algorithm~\ref{alg-change-neigh}),
$uR_0'L_0'=u$ satisfying Constraint~\ref{c-right-left}.
In the absence of a 
successor, $u$ simply sets $uR_0'=\nil$ (line~\ref{in2} 
or~\ref{in26} of Algorithm~\ref{alg-insert-new-node}),
thus trivially satisfying Constraints~\ref{c-right-bigger}
and~\ref{c-right-left}.

Node $u$ uses its neighbors at level $\ell$, $(\ell\geq0)$, to find its 
neighbors at level $\ell+1$. Node $u$ sends a \op{buddyOp} message to
$uR_{\ell}>u$ (line~\ref{in32} of Algorithm~\ref{alg-insert-new-node}),
and this message is passed to the right to successive nodes 
$uR_{\ell}^k>u$, $k\geq 1$, until it reaches a node $y$ such that 
$m(u)\restrictedto ({\ell}+1) = m(y)\restrictedto ({\ell}+1)$, and 
$u$ sets $uR_{{\ell}+1}'=y>u$ (line~\ref{in34} of 
Algorithm~\ref{alg-insert-new-node}), satisfying 
Constraint~\ref{c-right-bigger}.  As this message is only
sent to the right, $u$ can only connect on its right to nodes
greater than itself. As $y$ sets $yL_{\ell+1}'=u<y$, 
$uR_{\ell+1}'L_{\ell+1}'=u$, satisfying Constraint~\ref{c-right-left}.
Similarly, $u$ also sends a \op{buddyOp} messages to its
left to $uL_{\ell}<u$ (line~\ref{in37} of Algorithm~\ref{alg-insert-new-node}); 
as this message is only sent to nodes $s$ less that $u$, it ensures that 
$sR_{\ell+1}'=u>s$, satisfying Constraint~\ref{c-right-bigger}.
As $u$ sets $uL_{\ell+1}'=s$, $sR_{\ell+1}'L_{\ell+1}'=s$ 
satisfying Constraint~\ref{c-right-left}.

As $u$ only queries the nodes $z$ in the same list as itself 
at level $\ell$, it is ensured that 
$m(u) \restrictedto \ell = m(z)\restrictedto \ell$.
Further, we see that $u$ only links to a node $z$ such 
that $m(u)_{\ell} = m(z)_{\ell}$ (line~\ref{io10} of 
Algorithm~\ref{alg-insert-old-node}). 
Thus $u$ can only link to $z$ at level $({\ell}+1)$ if
$m(u)\restrictedto ({\ell}+1) = m(z)\restrictedto ({\ell}+1)$.
Having seen that $u$ only links to nodes with the
correct membership vector prefix, it only remains to show that 
$u$ links to the nearest such nodes at each level ${\ell} > 0$. We
see that $u$ starts looking for it successor at level ${\ell+1}$
from $uR_{\ell}$ (line~\ref{in31} of Algorithm~\ref{alg-insert-new-node}).
Node $uR_{\ell}$ is either the successor for node $u$ at level
$\ell+1$ (line~\ref{io10} of Algorithm~\ref{alg-insert-old-node}),
or it passes the message to its successor
(line~\ref{in14} of Algorithm~\ref{alg-insert-old-node}).
As the search for $uR_{\ell+1}$ proceeds one node at a time along 
$S_{m(u)\restrictedto (\ell)}$, it is guaranteed to find the
{\em nearest} node greater than $u$, whose membership vector matches 
$m(u)_{\ell}$. Thus $uR_{\ell+1} = uR_{\ell}^k$, for the smallest 
$k>0$, satisfying Constraint~\ref{c-right-successor}. 

We note that with concurrent inserts, additional nodes may get
linked at some level between $u$ and its predetermined neighbors,
found using either the search operation (for level $0$) or the
\op{buddyOp} messages (for levels greater than $0$). In each case, 
we see that when some old node receives a \op{getlinkOp} messages 
to link to $u$, it verifies that pointing to $u$ will maintain
the skip graph node order. Otherwise, it passes the message
to its appropriate neighbor (line~\ref{cn4} of 
Algorithm~\ref{alg-change-neigh}). This is explained in detail
above, and it ensures that $u$ links to the correct nodes at each 
level. So the constraints are maintained even with concurrent inserts 
in the skip graph.

Thus when all the concurrent insert operations are completed,
we get a skip graph. 
\end{proof}


\begin{lemma}
Deleting node $u$ from a skip graph $S$ using 
Algorithm~\ref{alg-delete-new} gives a skip graph.
\end{lemma}
\begin{proof}
Deleting a node $u$  from a skip graph $S$ consists of two stages 
at each level $\ell$: finding a node to the right of $u$ that
is not being deleted, and then finding a node to the left of
$u$ that is not being deleted to link these two nodes together.

Node $u$ sends a \op{deleteOp} message to its successor $uR_{\ell}$
(line~\ref{dn4} of Algorithm~\ref{alg-delete-new}). As long as
this message is received by a node that is itself being deleted
(line~\ref{do2} of Algorithm~\ref{alg-delete-other}), it is passed 
on to the right to successive nodes $uR_{\ell}^2, uR_{\ell}^3\ldots$
(line~\ref{do4} of Algorithm~\ref{alg-delete-other}), until it reaches
some node $z>u$ which is not being deleted. Node $z$ sends a 
\op{findNeighborOp} message to $zL_{\ell}$ to determine its new left 
neighbor (line~\ref{do8} of Algorithm~\ref{alg-delete-other}).
As long as
this message is received by a node that is itself being deleted
(line~\ref{do13} of Algorithm~\ref{alg-delete-other}), it is passed 
on to the left to successive nodes $zL_{\ell}^2, zL_{\ell}^3\ldots$
(line~\ref{do15} of Algorithm~\ref{alg-delete-other}), until it reaches
some node $s<u<z$ which is not being deleted. Node $s$ sends a
\op{foundNeighborOp} message back to $z$ and sets $sR_{\ell}'=z>s$, 
satisfying Constraint~\ref{c-right-bigger} (lines~\ref{do19} 
and~\ref{do20} of Algorithm~\ref{alg-delete-other}). Upon receipt
of this message (line~\ref{do9} of Algorithm~\ref{alg-delete-other}),
$z$ sets $zL_{\ell}'=s<z$ (line~\ref{do10} of Algorithm~\ref{alg-delete-other}), 
satisfying Constraint~\ref{c-left-smaller}.
Further, as $zL_{\ell}'R_{\ell}'=z$ and $sR_{\ell}'L_{\ell}'=s$,
Constraints~\ref{c-right-left} and~\ref{c-left-right} are also
satisfied. Nodes $z$ and $s$ are in the same list at level $\ell$,
so $m(z)\restrictedto \ell = m(s)\restrictedto \ell$, and they are
also in the same list at level $\ell-1$. As all the nodes between 
them will be eventually deleted, they are nearest nodes with 
matching membership vectors to be linked at level $\ell$,
satisfying Constraints~\ref{c-right-successor} 
and~\ref{c-left-successor}.

If no suitable node $s$ exists, the last node on the left of $z$
that is being deleted, sends a \op{foundNeighborOp} to $z$
(line~\ref{do17} of Algorithm~\ref{alg-delete-other}). Node
$z$ sets $zL_{\ell}'=\nil$ (line~\ref{do10} of 
Algorithm~\ref{alg-delete-other}), thus trivially satisfying 
Constraints~\ref{c-left-smaller},~\ref{c-left-right} 
and~\ref{c-left-successor}. If no suitable node $z$ exists,
the last node to the right of $u$ that is being deleted,
informs $u$ of that (line~\ref{do6} of Algorithm~\ref{alg-delete-other}).
Node $u$ sends a \op{setNeighborNilOp} message to $uL_{\ell}$
(line~\ref{dn8} of Algorithm~\ref{alg-delete-new}).
which is passed to the left until it reaches a node $q$ that 
is not being deleted (line~\ref{do24} of Algorithm~\ref{alg-delete-other}). 
Node $q$ sets $qR_{\ell}'=\nil$ (line~\ref{do29} of
Algorithm~\ref{alg-delete-new}), once again trivially 
satisfying Constraints~\ref{c-right-bigger},~\ref{c-right-left}
and~\ref{c-right-successor}. If no suitable node $q$ exists,
then no link changes are made at all, and all the constraints
are satisfied as before.

We note that with concurrent deletes, additional nodes may get
deleted at some level between $u$ and its existing neighbors.
We see that when some neighboring node receives a message for
the delete operation, it verifies that it is not being deleted
itself. If so, it passes the message on to its neighbor as
explained in detail above. This ensures that only nodes on
either side of $u$ that are not being deleted link to each
other. 

Thus, after all the concurrent delete operations have been completed, 
we get a skip graph.
\end{proof}


\section{Fault tolerance}
\label{section-fault-tolerance}

In this section, we describe some of the fault tolerance properties 
of a skip graph with alphabet $\{0,1\}$.
Fault tolerance of related data structures, such as
augmented versions of linked lists and binary trees, has been
well-studied and some results can be seen in~\cite{MunroP1984,
AumannB1996}.  In Section~\ref{section-repair}, we give a 
repair mechanism that detects node failures and initiates
actions to repair these failures. Before we explain the repair
mechanism, we are interested in the number of nodes that can be 
separated from the primary component by the failure of other nodes, 
as this determines the size of the surviving skip graph after the
repair mechanism finishes. 

Note that if multiple nodes are stored on a single machine, when that 
machine crashes, all of its nodes vanish simultaneously. Our results
are stated in terms of the fraction of nodes that are lost; if the nodes
are roughly balanced across machines, this will be proportional to the
fraction of machine failures. Nonetheless, it would be useful to have a 
better understanding of fault tolerance when the mapping of resources
to machines is taken into account; this may in fact dramatically
improve fault tolerance, as nodes stored on surviving machines can always
find other nodes stored on the same machine, and so need not be lost even 
if all of their neighbors in the skip graph are lost.

We consider two fault models: a random failure model in which an adversary 
chooses random nodes to fail, and a worst-case failure model in which an 
adversary chooses specific nodes to fail after observing the structure 
of the skip graph
For a random failure pattern, experimental results, presented in
Section~\ref{section-faults-experiments}, show that for a reasonably
large skip graph nearly all nodes remain in the primary component
until about two-thirds of the nodes fail, and that it is possible to
make searches highly resilient to failures even without using the
repair mechanism, by the use of redundant links.
For a worst-case failure pattern, theoretical results, presented in
Section~\ref{section-faults-expansion}, show that even a worst-case choice 
of failures causes limited damage. With high probability, a skip graph with 
$n$ nodes has an $\Omega(\frac{1}{\log n})$ expansion ratio, implying that 
at most $O(f\cdot \log n)$ nodes can be separated from the primary component 
by $f$ failures. We do not give experimental results for adversarial
failures as experiments may not be able to identify the worst-case
failure pattern.


\subsection{Random failures}
\label{section-faults-experiments}

\begin{figure}[htp]
  \begin{center}
  \includegraphics[width=240pt]{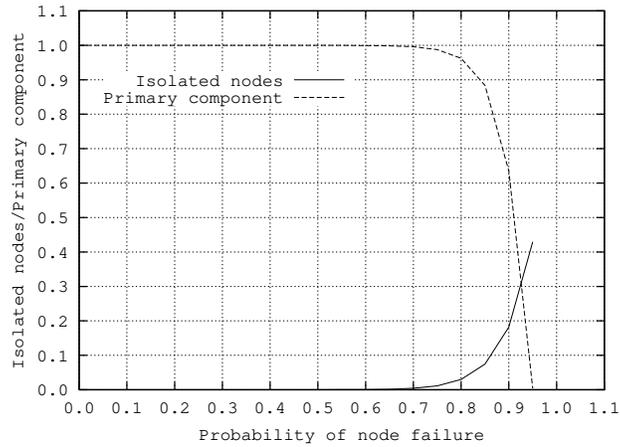}
  \caption[Isolated nodes/Primary component with node failures.]
  {The number of isolated nodes and the size of the primary
   component as a fraction of the
   surviving nodes in a skip graph with $131072$ nodes.}
  \label{figure-random-failures}
  \end{center}
\end{figure}

In our simulations, skip graphs appear to be highly resilient to random 
failures. We constructed a skip graph of $131072$ nodes, where each node
was had a unique label from $[1,131072]$. We progressively
increased the probability of node failure and measured the size
of largest connected component of the live nodes as well as the
number of isolated nodes as a fraction
of the total number of nodes in the graph.  As shown in
Figure~\ref{figure-random-failures}, nearly all nodes remain in the
primary component even as the probability of individual node failure
exceeds $0.6$. We also see that a lot of nodes are isolated as the 
failure probability increases because all of their immediate neighbors die.

\begin{figure}[htp]
  \begin{center}
  \includegraphics[width=240pt]{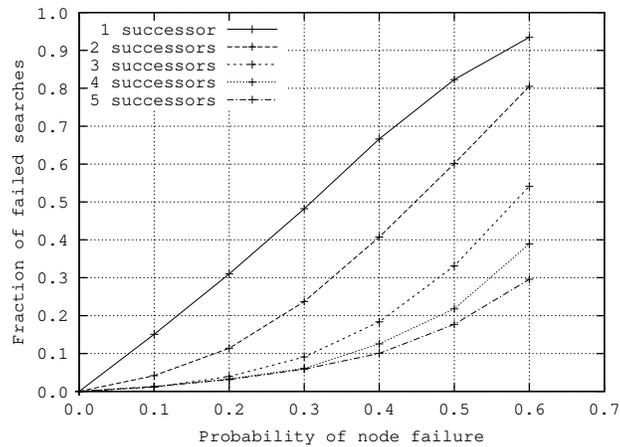}
  \caption[Fraction of failed searches with failed nodes.]
  {Fraction of failed searches in a skip graph with 
   $131072$ nodes and $10000$ messages. Each node has up to
   five successors at each level.
   }
  \label{figure-search-results}
  \end{center}
\end{figure}
For searches,
the fact that the average search involves only $O(\log n)$ nodes
establishes trivially that most searches succeed as long as the
proportion of failed nodes is substantially less than 
$O(\frac{1}{\log n})$.
By detecting failures locally and using
additional redundant edges, we can make searches highly tolerant to
small numbers of random faults.

Some further experimental results are shown
in Figure~\ref{figure-search-results}. In these experiments, each
node had additional links to up to five nearest successors
at every level. A total of $10000$ messages were sent
between randomly chosen source and destination nodes, and the
fraction of failed searches was measured. We see that skip graphs
are quite resilient to random failures. This plot appears
to contradict the one shown in Figure~\ref{figure-random-failures},
because we would expect all the searches to succeed as long as all
live nodes are in the same connected component. However, once
the source and target nodes are fixed, there is a fixed, deterministic
path along which the search proceeds and if any node on this
path fails, the search fails. So there may be {\em some} path
between the source and the destination nodes, putting
them in the same connected component, but the path 
used by the search algorithm may be broken, foiling the search. This 
suggests that if we use smarter search techniques, such as 
jumping between the different skip lists that a node belongs to,
we can get much better search performance even in the presence 
of failures.

In general, skip graphs do not provide as strong guarantees as those 
provided by data structures based on explicit use of expanders such as
censorship-resistant networks~\cite{FiatS2002, SaiaFGKS2002, Datar2002}. 
But we believe that this is compensated for by the simplicity of
skip graphs and the existence of good distributed
mechanisms for constructing and repairing them.


\subsection{Adversarial failures}
\label{section-faults-expansion}

In addition to considering random failures, we are also interested
in analyzing the performance 
of a skip graph when an adversary can observe the data structure, and 
choose specific nodes to fail. Experimental results may not 
even be able to identify these worst-case failure patterns. So in this 
section, we look at the expansion ratio of a skip graph, as that gives 
us the number of nodes that can be separated from the primary component 
even with adversarial failures.

Let $G$ be a graph. Recall that the expansion ratio of a set of nodes
$A$ in $G$ is $|\delta A|/|A|$, where $|\delta A|$ is the number of
nodes that are not in $A$ but are adjacent to some node in $A$. The
expansion ratio of the graph $G$ is the minimum expansion ratio for 
any set $A$, for which $1 \le |A| \le n/2$.
The expansion ratio determines the resilience of a graph in the
presence of adversarial failures, because separating a set $A$ from
the primary component requires all nodes in $\delta A$ to fail.
We will show that skip graphs have $\Omega(\frac{1}{\log n})$ expansion ratio
with high probability, implying that only $O(f \cdot \log n)$ nodes can be
separated by $f$ failures, even if the failures are carefully
targeted.

Our strategy for showing a lower bound on the
expansion ratio of a skip graph will be to show that with high
probability, all sets $A$ either have large $\delta_0 A$ (i.e., many
neighbors at the bottom level of the skip graph) or have large
$\delta_{\ell} A$ for some particular $\ell$ chosen based on the size of $A$.
Formally, we define $\delta_{\ell} A$ as the set of all nodes that are not in
$A$ but are joined to a node in $A$ by an edge at level ${\ell}$.  Our result
is based on the observation that $\delta A = \bigcup_{\ell} \delta_{\ell} A$ and
$|\delta A| \ge \max_{\ell} |\delta_{\ell} A|$.
We begin by counting the number of sets $A$ of a given size that have
small $\delta_0 A$.

\begin{lemma}
\label{lemma-delta-0}
In a $n$-node skip graph with alphabet $\{0,1\}$, 
the number of sets $A$, where $|A| = m < n$ and
$|\delta_0 A| < s$, is less than
$\LemmaDeltaZeroBound$.
\end{lemma}
\begin{proofsketch}
Represent each $A$ as a bit-vector where $1$ indicates a member of the
set and $0$ a non-member.  Then $|\delta_0 A|$ is at least the
number of intervals of zeroes in this bit-vector.  The bound in the
lemma is then obtained by bounding the number of length $n$
bit-vectors with $m$ ones
and at most $s$ intervals of zeroes.
\end{proofsketch}
\begin{proof}
Without loss of generality, assume that the nodes of the skip graph
are numbered from $1$ to $n$.  Given a subset $A$ of these nodes,
define a corresponding bit-vector $x$ by letting $x_i = 1$ if and only
if node $i$ is in $A$.  Then $\delta_0 A$ corresponds to all zeroes
in $x$ that are adjacent to a one.

Consider the extended bit-vector $x' = 1x1$ obtained by appending a
one to each end of $x$.
Because $x'$ starts and ends with a one, it can be divided into
alternating intervals of ones and zeroes, of which $r+1$ intervals
will consist of ones and $r$ will consist of zeroes for some $r$,
where $r > 0$ since $x$ contains at least one zero.
Observe that each interval of zeroes contributes at least one and at
most two of its endpoints to $\delta_0 A$.  It follows that
$r \le |\delta_0 A| \le 2r$, and thus any $A$ for which $|\delta_0 A| <
s$ corresponds to an $x$ for which $x'$ contains $r \le |\delta_0 A| < s$
intervals of zeroes.

Since there is at least one $A$ with $r < s$ but $|\delta_0 A| \ge s$,
the number of sets $A$ with $|\delta_0 A| < s$ is strictly less than the
number of sets $A$ with $r < s$.  By counting the latter quantity, we
get a strict upper bound on the former.

We now count, for each $r$, the number of bit-vectors $x'$ with $n-m$
zeroes consisting of
$r+1$ intervals of ones and $r$ intervals of zeroes.
Observe that we can characterize such a bit-vector completely by
specifying the nonzero length of each of the $r+1$ all-one intervals
together with the nonzero length of each of the $r$ all-zero intervals.
There are $m+2$ ones that must be distributed among the $r+1$ all-one
intervals, and there are ${m+2-1 \choose r+1-1}={m+1 \choose r}$ ways
to do so.  Similarly, there are $n-m$ zeroes to distribute among the
$r$ all-zero intervals, and there are ${n-m-1\choose r-1}$ ways to do so.
Since these two distributions are independent, the total count is
exactly ${m+1 \choose r}{n-m-1 \choose r-1}$.

Summing over all $r < s$ then gives the upper bound
$\sum_{r=1}^{s-1} {m+1 \choose r}{n-m-1 \choose r-1}$.
\end{proof}


For levels $\ell>0$, we show with a probabilistic argument,
that $|\delta_{\ell} A|$ is only rarely small


\newcommand{\LemmaDeltaHBound}{2 {2^{\ell} \choose \LemmaDeltaHFloorThing} (2/3)^{m}}
\begin{lemma}
\label{lemma-delta-h}
Let $A$ be a subset of $m \le n/2$ nodes of a $n$-node skip graph $S$
with alphabet $\{0,1\}$.
Then for any $\ell$,
$\LemmaDeltaHLHS < \LemmaDeltaHBound$.
\end{lemma}
\begin{proof}
\newcommand{\notA}{S-A}
The key observation is that for each $b$ in $\{0,1\}^{\ell}$,
if $A$ contains a node $u$ with $m(u)\restrictedto \ell = b$
and $A$'s complement $\notA$
contains a node $v$ with $m(v)\restrictedto \ell = b$,
then there exist nodes $u' \in A$ and $v' \in \notA$
along the path from $u$ to $v$ in $S_b$,
such that $u'$ and $v'$ are adjacent in $S_b$.
Furthermore, since such pairs are distinct for distinct $b$,
we get a lower bound on $\delta_{\ell} A$
by computing a lower bound on the number of
distinct $b$ for which $A$ and $\notA$ both contain
at least one node in $S_b$.

Let $T(A)$ be the set of $b \in \{0,1\}^{\ell}$ for which $A$ contains a
node of $S_b$, and similarly for $T(\notA)$.
Then
\begin{eqnarray*}
\Pr\left[|T(A)| < \frac{2}{3}\cdot 2^{\ell}\right]
&\le&
\sum_{B \subset S, |B| = \LemmaDeltaHFloorThing}
 \Pr\left[T(A) \subseteq B\right]
\\
&\le&
{2^{\ell} \choose \LemmaDeltaHFloorThing} (2/3)^{|A|},
\end{eqnarray*}
and by the same reasoning,
\begin{eqnarray*}
\Pr\left[|T(\notA)| < \frac{2}{3}\cdot 2^{\ell}\right]
&\le&
{2^{\ell} \choose \LemmaDeltaHFloorThing} (2/3)^{|\notA|}.
\end{eqnarray*}
But if both $T(A)$ and $T(\notA)$ hit at least two-thirds of the 
$b$, then their intersection must hit at least one-third,
and thus
the probability that $T(A) \cap T(\notA) < \frac{1}{3}\cdot 2^{\ell}$
is at most
${2^{\ell} \choose \LemmaDeltaHFloorThing}
 \left((2/3)^{|A|} + (2/3)^{|\notA|}\right)$,
which is in turn bounded by
$2 {2^{\ell} \choose \LemmaDeltaHFloorThing} (2/3)^{|A|}$
under the assumption that $|A| \le |\notA|$.
\end{proof}


We can now get the full result by arguing that there are not enough
sets $A$ with small $|\delta_0 A|$ (Lemma~\ref{lemma-delta-0})
to get a non-negligible probability that any one of them has small
$|\delta_{\ell} A|$ for an appropriately chosen ${\ell}$ (Lemma~\ref{lemma-delta-h}).
Details are given in the proof of Theorem~\ref{theorem-expansion} below.


\newcommand{\ltt}{\lg(3/2)}
\newcommand{\lgt}{\log_{3/2}}
\begin{theorem}
\label{theorem-expansion}
Let $c \ge 6$.
Then a skip graph with $n$ nodes and alphabet $\{0,1\}$,
has an expansion ratio of at least $\frac{1}{c \lgt n}$
with probability at least $1-\alpha n^{5-c}$,
where the constant factor $\alpha$ does not depend on $c$.
\end{theorem}
\begin{proofsketch}
The probability bound is obtained by summing the probability of having
$\delta_{\ell} A$ too small over all $A$
for which $\delta_0 A$ is too small.  For each set $A$ of size $m$,
${\ell}$ is chosen
so that the $\frac{1}{3}\cdot 2^{\ell}$ bound of Lemma~\ref{lemma-delta-h}
exceeds $m$ times the expansion ratio.
The probabilities derived from Lemma~\ref{lemma-delta-h} are then
summed over all sets $A$ of a fixed size $m$ using Lemma~\ref{lemma-delta-0},
and the result of this process is
summed over all $m > c \lgt n$ to obtain the final bound.
\end{proofsketch}
\begin{proof}
We will show that the probability that a skip graph $S$
with $n$ nodes
does \emph{not} have the given
expansion ratio is at most $\alpha n^{5-c}$, where $\alpha=31$.
The particular value of $\alpha=31$ may be an artifact of our
proof; the actual constant may be smaller.

Consider some subset $A$ of $S$ with $|A| = m \le n/2$.
Let $s = \frac{m}{c \lgt n} = \frac{m \ltt}{c \lg n}$,
and let $s_1 = \left\lceil s \right\rceil$.
We wish to show that, with high probability, all $A$ of size $m$ have
$|\delta A| \ge s_1$.
We will do so by counting the expected number of sets $A$ with smaller
expansion.  Any such set must have both $|\delta_0 A| < s$ and
$|\delta_{\ell} A| < s$ for any ${\ell}$; our strategy will be to show first
that there are few sets $A$ in the first category, and that each
set that does have small $\delta_0 A$ is very likely to have large
$\delta_{\ell} A$ for a suitable ${\ell}$.

By Lemma~\ref{lemma-delta-0}, there are at most
$\LemmaDeltaZeroBoundOne$ sets $A$ of size $m$ for which
$|\delta_0 A| < s_1$.
It is not hard to show that the largest term in this sum dominates.
Indeed, for $r < s_1$, the ratio between adjacent terms
${m+1 \choose r}{n-m-1 \choose r-1}
 / {m+1 \choose r+1}{n-m-1 \choose r}$
equals $\frac{r}{m+1-r} \cdot \frac{r+1}{n-m-r}$,
and since $r<\frac{1}{6}m$ and $m\leq \frac{n}{2}$,
this product is easily seen to be less than $\frac{1}{2}$. 
It follows that
\begin{eqnarray*}
\LemmaDeltaZeroBoundOne &<& {m+1 \choose s_1-1}{n-m-1 \choose s_1-2}
\sum_{i=0}^{\infty} 2^{-i}
\\
&=& 2 {m+1 \choose s_1-1}{n-m-1 \choose s_1-2}
\\
&<& n^{2(s_1-1)}
\\
&<& n^{2s}
= n^{\frac{2m}{c \lgt n}}
= 2^{\frac{2m \ltt}{c}}.
\end{eqnarray*}

Now let ${\ell} = \lceil \lg s + \lg 3\rceil$, so that $s \le
\frac{1}{3} 2^{\ell} \le 2s$. 

Applying Lemma~\ref{lemma-delta-h},
we have
\begin{eqnarray*}
\Pr\left[|\delta_{\ell} A| < s|\right]
&\le&
\LemmaDeltaHLHS
\\
&<&
\LemmaDeltaHBound
\\
&\le&
2 {6s \choose 2s+1} (2/3)^m
\\
&<&
2 \cdot (6s)^{2s+1} (2/3)^m,
\end{eqnarray*}
and thus
\begin{equation}
\label{theorem-expansion-ugly-bound}
\sum_{A \subset S, |A| = m, |\delta_0 A| < s}
 \Pr\left[|\delta_{\ell} A| < s|\right]
<
2^{\frac{2m \ltt}{c}}\cdot 2 \cdot (6s)^{2s+1} (2/3)^m.
\end{equation}

Taking the base-$2$ logarithm of the right-hand side gives
\begin{eqnarray*}
\lefteqn{\frac{2m\ltt}{c} + 1 + (2s+1) \lg(6s) - m \ltt}
\\
&<&
\frac{2m\ltt}{c} + 1 + \left(\frac{3m\ltt}{c \lg n}\right) \lg n - m \ltt
\\
&=&
1+m\left(\frac{5}{c} - 1\right) \ltt.
\end{eqnarray*}
It follows that the probability that there exists an $A$ of size $m$,
for which both $|\delta_0 A|$ and $|\delta_{\ell} A|$ are less than $s$, is
at most $2$ to the above quantity, which we can write as $2b^m$ where
$b = (3/2)^{(5/c-1)}$.

To compute the probability that any set has a small neighborhood, we
sum over $m$.  By definition of the expansion ratio, we need only
consider values of $m$ less than or equal to $n/2$; however, because
every proper subset $A$ of $S$ has at least one neighbor, we need
to consider only $m > c \lgt n$.  So we have
\begin{eqnarray*}
\Pr\left[\mbox{$S$ has expansion ratio less than $\frac{1}{c \lgt n}$}\right]
&<&
\sum_{m=\lceil c \lgt n \rceil}^{n/2} 2b^m
\\
&<&
2\left( \sum_{i=0}^{\infty}b^i\right) \left( b^{\lceil c \lgt n \rceil}\right)
\\
&\leq&
\frac{2}{1-b} \cdot b^{c \lgt n}
\\
&=&
\frac{2}{1-b} \cdot n^{c \lgt b}
\\
&=&
\frac{2}{1-b} \cdot n^{c \left(\frac{5}{c}-1\right)}
\\
&=&
\frac{2}{1-b} \cdot n^{5-c}
\\
&<& 31 \cdot n^{5-c},
\end{eqnarray*}
where the last inequality follows from the assumption that $c \ge 6$.
\end{proof}


\section{Repair mechanism}
\label{section-repair}

Although  a skip graph can survive a few disruptions, it
is desirable to avoid accumulating errors. A large number of 
unrepaired failures will degrade the ideal search performance 
of a skip graph. Replication alone may not guarantee robustness,
and we need a repair mechanism that automatically heals
disruptions. Further, as failures occur continuously, the
repair mechanism needs to continuously monitor the state of
the skip graph to detect and repair these failures. The goal of 
the repair mechanism is to take a defective skip graph and repair 
all the defects.

We describe the repair mechanism as follows: In 
Section~\ref{section-maintain-invariant}, we show that the first two
skip graph constraints, given in section~\ref{section-correctness},
are always preserved in any execution of the skip graph. In 
Section~\ref{section-repair-constraints}, we show how the 
remaining constraints can be checked locally by every node, and 
give algorithms to repair errors which may exist in the data structure. 
In Section~\ref{section-put-together}, we prove that the repair
mechanism given in Section~\ref{section-repair-constraints}
repairs a defective skip graph to give a defectless one.
We note that the repair mechanism is
not a \buzz{self-stabilization} mechanism in the strong sense because
it will not repair an arbitrarily linked skip graph and
restore it to its valid state. Instead, we see that certain 
defective configurations are impossible given the particular types
of failures we consider. Thus the repair mechanism only repairs
those failures that occur starting from a defectless skip graph.


\subsection{Maintaining the invariant}
\label{section-maintain-invariant}

We again list the constraints that describe a skip graph, as given
in Section~\ref{section-correctness}.
Let $x$ be any node in the skip graph; then for all levels $\ell\geq 0$:

\begin{enumerate}
  \item If $xR_{\ell} \neq \nil$, $xR_{\ell} > x$.

  \item If $xL_{\ell} \neq \nil$, $xL_{\ell} < x$.

  \item If $xR_{\ell} \neq \nil$, $xR_{\ell}L_{\ell} = x$.

  \item If $xL_{\ell} \neq \nil$, $xL_{\ell}R_{\ell} = x$.

  \item If $m(x)\restrictedto (\ell+1) = m(xR_{\ell}^{k}) \restrictedto (\ell+1)$
        and $\nexists j, j<k, m(x)\restrictedto (\ell+1) = m(xR_{\ell}^j) 
        \restrictedto (\ell+1)$, then $xR_{\ell+1} = xR_{\ell}^{k}$. Else,
        $xR_{\ell+1}=\nil$.

  \item If $m(x)\restrictedto (\ell+1) = m(xL_{\ell}^{k}) \restrictedto (\ell+1)$
        and $\nexists j, j<k, m(x)\restrictedto (\ell+1) = m(xL_{\ell}^j) 
        \restrictedto (\ell+1)$, then $xL_{\ell+1} = xL_{\ell}^{k}$. Else,
        $xL_{\ell+1}=\nil$.
\end{enumerate}

We define Constraints~\ref{c-right-bigger} and~\ref{c-left-smaller}
as an \buzz{invariant} for a skip graph as they hold in all states 
even in the presence of failures. Constraints~\ref{c-right-left}
to~\ref{c-left-successor} may fail to hold with failures, but they 
can be restored by the repair mechanism. We call 
Constraints~\ref{c-right-left} and~\ref{c-left-right} the
$R$ and $L$ \buzz{backpointer} constraints respectively, and  
Constraints~\ref{c-right-successor} and~\ref{c-left-successor}
the $R$ and $L$ \buzz{inter-level} constraints respectively.
Each node periodically checks to see if its backpointer or
inter-level constraints have been violated. If it discovers
an inconsistent constraint, it initiates the repair mechanism 
explained in Section~\ref{section-repair-constraints}.

We consider the failure of some node $x$ as an atomic action which 
eliminates $x$ from the skip graph, and effectively sets the
corresponding pointers to it to $\nil$. If there are any pending 
messages to other nodes for them to change their pointers to
point to $x$, when the messages are delivered, the corresponding
pointers are set to $\nil$. In an actual implementation, each
node $y$ will periodically check to see if its neighbors at each
level $\ell$ are alive, and in absence of a response, it will set 
$yR_{\ell}=\nil$ or $yL_{\ell}=\nil$. For the purposes of our
proofs however, we will consider that when a node fails, the pointers
to it are atomically set to $\nil$. It is possible that a node
will detect its failed neighbors before it has to initiate some
action, thus setting the pointers to $\nil$ anyway. If it does not, 
we can ensure that a node checks that its
neighbors at level $\ell$ are alive before it processes a 
message that it receives for level $\ell$. 

We use $xL_{\ell}'$ and $xR_{\ell}'$ etc to denote the value of  
node $x$'s predecessor and successor respectively {\em after}
some operation has occurred.

We prove that the invariant is maintained for the insert and delete 
operations in the presence of node failures. Then we give a repair 
mechanism that uses the invariant constraints to repair any violated 
backpointer or inter-level constraints due to node failures.

\begin{lemma}
\label{lemma-fail}
Failures preserve the invariant when no operation is in progress.
\end{lemma}
\begin{proof}
Suppose that the invariant holds in the absence of any 
failures. If a node $x$ has a failed successor or predecessor 
at level $i$, we consider $xR_i'=\nil$ or $xL_i'=\nil$ respectively, 
which trivially satisfies Constraints~\ref{c-right-bigger} 
and~\ref{c-left-smaller}. Thus, for all nodes $y$ and all levels 
$\ell$, $yR_{\ell}'$ and $yL_{\ell}'$ are either equal to their 
previous values or they are set to $\nil$, and the invariant is 
maintained.
\end{proof}

\begin{lemma}
\label{lemma-insert}
The invariant is maintained during an insert operation even
in the presence of failures.
\end{lemma}
\begin{proof}
Suppose that the invariant holds prior to the insert operation.
We consider each link change during an insert operation and prove that 
this does not violate Constraint~\ref{c-right-bigger}; we omit the
details for Constraint~\ref{c-left-smaller} as it is a mirror image
of Constraint~\ref{c-left-smaller}. We also consider only the case where
the introducing node $s$ is less than the new node $u$ that is being 
added as the other case is similar.  The successor link changes in 
Algorithm~\ref{alg-insert-new-node} (and its subroutine 
Algorithm~\ref{alg-change-neigh}) during the insert operation are as 
follows:
  \begin{itemize}
    \item Line~\ref{in23}:  Node $u$ sends a \op{getLinkOp}
          message to $s$ (line~\ref{in21} of Algorithm~\ref{alg-insert-new-node}). 
          Node $s$ either
          returns a \op{setLinkOp} message to $u$ (line~\ref{cn6} of
          Algorithm~\ref{alg-change-neigh}), or passes the message to
          $sL_0$ (line~\ref{cn4} of Algorithm~\ref{alg-change-neigh})
          if $u<s$. The
          latter case occurs when a new node has been inserted between 
          $s$ and $sL_0$ during concurrent inserts. Thus $u$'s original
          \op{getLinkOp} messages is only passed to nodes smaller than 
          $u$, and $u$ receives the corresponding \op{setLinkOp} from a 
          node $p$ smaller than itself. Thus $pR_0'=u>p$ (line~\ref{cn7}
          of Algorithm~\ref{alg-change-neigh}). If some node
          fails in this process or no suitable predecessor exists, all $R$ links
          remain unchanged, thus satisfying Constraint~\ref{c-right-bigger}.
    \item Line~\ref{in26}:  Node $u$ sends a \op{getLinkOp}
          message to $sR_0>u$ (line~\ref{in21} of Algorithm~\ref{alg-insert-new-node}). 
          Node $sR_0$ either
          returns a \op{setLinkOp} message to $u$ (line~\ref{cn6} of
          Algorithm~\ref{alg-change-neigh}), or passes the message to
          $sR_0^2$ (line~\ref{cn4} of Algorithm~\ref{alg-change-neigh})
          if $sR_0<u$. The
          latter case occurs when a new node has been inserted between 
          $s$ and $sR_0$ during concurrent inserts. Thus $u$'s original
          \op{getLinkOp} messages is only passed to nodes greater than 
          itself, and it receives the corresponding \op{setLinkOp} from a 
          node $v$ greater than itself. Thus $uR_0'=v>u$. If some node
          fails in this process or no suitable successor exists, 
          $uR_0'=\nil$, thus satisfying Constraint~\ref{c-right-bigger}.
    \item Line~\ref{in34}: For level $\ell\geq 0$, $u$ sends a \op{buddyOp}
          message to $uR_{\ell}>u$ (line~\ref{in32} of Algorithm~\ref{alg-insert-new-node}). 
          If $m(uR_{\ell})_{\ell}=m(u)_{\ell}$, then $uR_{\ell}$ sends a
          \op{setLinkOp} message to $u$ (line~\ref{cn6} of 
          Algorithm~\ref{alg-change-neigh}), and $uR_{\ell+1}'=uR_{\ell}>u$.
          Else, $uR_{\ell}$ sends the \op{buddyOp} message to $uR_{\ell}^2$
          (line~\ref{cn4} of Algorithm~\ref{alg-change-neigh}).
          Node $u$'s original \op{buddyOp} message is only sent to nodes
          greater than itself, and it receives the corresponding \op{setLinkOp} 
          message only from a node $z$ greater than itself. Thus $uR_{\ell+1}'=z>u$.
          If some node fails in this process or no suitable successor exists, 
          $uR_{\ell+1}'=\nil$, thus satisfying Constraint~\ref{c-right-bigger}.
    \item Line~\ref{in39}: For level $\ell\geq 0$, $u$ sends a \op{buddyOp}
          message to $uL_{\ell}<u$ (line~\ref{in32} of Algorithm~\ref{alg-insert-new-node}). 
          If $m(uL_{\ell})_{\ell}=m(u)_{\ell}$, then $uL_{\ell}$ sends a
          \op{setLinkOp} message to $u$ (line~\ref{cn6} of Algorithm~\ref{alg-change-neigh}),
          and $uL_{\ell}R_{\ell+1}'=u>uL_{\ell}$.
          Else, $uL_{\ell}$ sends the \op{buddyOp} message to $uL_{\ell}^2$
          (line~\ref{cn4} of Algorithm~\ref{alg-change-neigh}).
          Node $u$'s original \op{buddyOp} message is only sent to nodes
          smaller than itself, and it receives the corresponding \op{setLinkOp} 
          message only from a node $s$ smaller than itself.  Thus $sR_{\ell+1}'=u<s$ 
          (line~\ref{cn7} of Algorithm~\ref{alg-change-neigh}). If some node fails 
          in this process or no suitable predecessor exists, all $R$ links remain 
          unchanged, thus satisfying Constraint~\ref{c-right-bigger}.
  \end{itemize}
Thus the invariant is maintained during an insert operation even in the
presence of failures.
\end{proof}

\begin{lemma}
\label{lemma-delete}
The invariant is maintained during a delete operation even
in the presence of failures.
\end{lemma}
\begin{proof}
Suppose that the invariant holds prior to the delete operation.
We consider each link change during a delete operation and prove that 
this does not violate Constraint~\ref{c-right-bigger}; we omit the
details for Constraint~\ref{c-left-smaller} as it is a mirror image
of Constraint~\ref{c-left-smaller}. The successor links changes in 
Algorithm~\ref{alg-delete-other} during the delete operation of node
$u$ are as follows:
  \begin{itemize}
     \item Line~\ref{do20}: At each level $\ell\geq0$, $u$ sends a 
           \op{deleteOp} message to 
           $uR_{\ell}$ (line~\ref{dn4} of Algorithm~\ref{alg-delete-new}).
           If $uR_{\ell}$ is being deleted, it passes the message to 
           $uR_{\ell}^2$ (line~\ref{do4} of Algorithm~\ref{alg-delete-other}). 
           This message is only passed
           to nodes greater than $u$ until it reaches a node $v>u$ which
           is not being deleted. Node $v$ sends a \op{findNeighborOp} 
           message to $vL_{\ell}$ (line~\ref{do8} of Algorithm~\ref{alg-delete-other}), 
           which is passed to 
           the left (line~\ref{do15}) of Algorithm~\ref{alg-delete-other} until 
           it reaches a node $s<u$ which
           is not being deleted. Node $s$ sends a \op{foundNeighborOp} to
           $v$ (line~\ref{do19} of Algorithm~\ref{alg-delete-other}), and it 
           sets $sR_{\ell}'=v>s$. If no 
           suitable $s$ is found (line~\ref{do17} of Algorithm~\ref{alg-delete-other}), 
           all the $R$ links
           remain unchanged, thus satisfying Constraint~\ref{c-right-bigger}.
     \item Line~\ref{do29}: If no suitable $v$, which is not being deleted, 
           is found (line~\ref{do6} of Algorithm~\ref{alg-delete-other}), $u$ sends 
           a \op{setNeighborNilOp} to 
           $uL_{\ell}<u$ (line~\ref{dn8} of Algorithm~\ref{alg-delete-other}). This 
           message is passed to the left 
           (line~\ref{do24} of Algorithm~\ref{alg-delete-other}) until it reaches some 
           node $s<u$ which is not
           being deleted. Node $s$ sets $sR_{\ell}'=\nil$ (line~\ref{do29} of 
           Algorithm~\ref{alg-delete-other}). 
           If no suitable $s$ is found (line~\ref{do17} of Algorithm~\ref{alg-delete-other}), 
           all the $R$ links
           remain unchanged, thus satisfying Constraint~\ref{c-right-bigger}.
  \end{itemize}
Thus the invariant is maintained during a delete operation even in the
presence of failures.
\end{proof}

Combining Lemmas~\ref{lemma-fail},~\ref{lemma-insert}
and~\ref{lemma-delete} directly gives Theorem~\ref{theorem-maintain}.

\begin{theorem}
\label{theorem-maintain}
The invariant is maintained throughout any execution of a skip graph, 
even with failures.
\end{theorem}


\subsection{Restoring invalid constraints}
\label{section-repair-constraints}

The backpointer and inter-level constraints are violated during insert and delete 
operations as well as when a node fails. However, we will see that the repair
mechanism needs to be triggered only for constraint violations caused due to failures, 
and not during the insert and delete operations. We give a repair mechanism in 
which each node periodically checks Constraints~\ref{c-right-left} to~\ref{c-left-successor} 
and initiates actions to fix invalid constraints due to node failures.

Although Constraints~\ref{c-right-left} to~\ref{c-left-successor} may be 
violated midway during an insert or a delete operation, once all the
pending operations are completed, these constraints are satisfied. Thus 
we observe that the repair mechanism is required to restore these constraints 
only in case of node failures. When a node fails during an insert or a delete, 
it leads to violations of the backpointer and inter-level constraints of its 
neighbors. Each node also periodically checks it backpointer and inter-level 
constraints. In Algorithm~\ref{alg-repair-backpointer}, node $x$ checks
its backpointer constraints by sending \op{checkNeighborOp} messages to 
its neighbors at all levels (lines~\ref{l3} and~\ref{l5}). Similarly,
each node checks its inter-level constraints as explained in 
Section~\ref{section-repair-successor}.

\begin{figure}[htp]
  \begin{center}
    \input{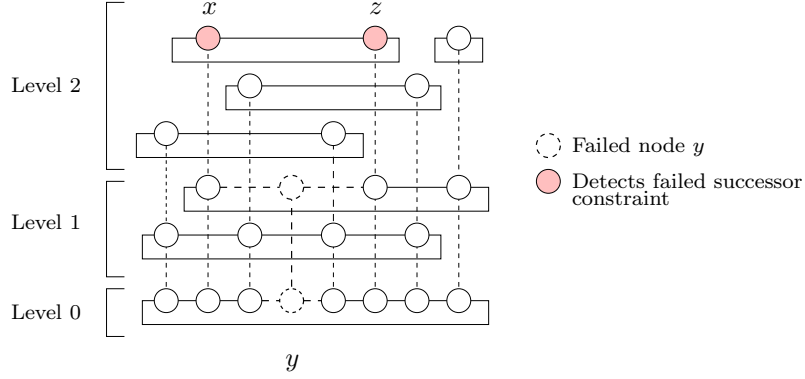}
    \caption[Violations of backpointer and inter-level constraints.]
    {Violation of backpointer and inter-level constraints when a node fails
     half-way through an insert or delete operation. Observe that
     $xR_2\neq x_1^j$, for any $j\geq 1$, and
     $zL_2\neq z_1^k$, for any $k\geq 1$.}
    \label{figure-fail-insert}
  \end{center}
\end{figure}

As shown in Figure~\ref{figure-fail-insert}, when node $y$ fails 
during an insert at level $2$ (after having successfully inserted itself at 
levels $0$ and $1$) or during a delete at level $1$ (after having successfully 
deleted itself from Level $2$), its neighbors $x$ and $z$ detect this failure. 
With the failure of $y$, $x$ and $z$ will detect inconsistencies in their constraints 
and initiate the mechanism to repair them. We prove that it is sufficient to detect 
and repair the violated constraints to restore the skip graph to its defectless state.

The repair mechanism is divided into two parts: the first part is used to
repair the the invalid backpointer constraints, and the second part is used 
to repair invalid inter-level constraints. 


\subsubsection{Restoring backpointer constraints}
\label{section-repair-link}

Each node $x$ periodically checks that $xR_{\ell}L_{\ell}=x$ when 
$xR_{\ell}\neq\nil$, and that $xL_{\ell}R_{\ell}=x$  when 
$xL_{\ell}\neq\nil$ for all levels $0\leq \ell\leq x.$maxLevel. It 
triggers the backpointer constraint repair mechanism 
(Algorithm~\ref{alg-repair-backpointer}) if it detects an inconsistency. 


\begin{algorithm}[htp]
\SetKwIf{gIf}{gElse}{else if}{then}{else}{}
\caption{Algorithm for repairing invalid backpointer constraints for node $v$.}
\label{alg-repair-backpointer}
  \lnl{l1}
  \For{$i\leftarrow v$.max\_levels \KwDownTo $0$}
  {
     \lnl{l2}
     \If{$(\neigh{v}{L}{i} \neq \nil)$}
     {
         \lnl{l3}
         \KwSend \msg{checkNeighborOp}{$R$, $v$, $i$} to \neigh{v}{L}{i}\;
     }
     \lnl{l4}
     \If{$(\neigh{v}{R}{i} \neq \nil)$}
     {
         \lnl{l5}
         \KwSend \msg{checkNeighborOp}{$L$, $v$, $i$} to \neigh{v}{R}{i}\;
     }
  }
  \vspace*{5pt}
  \lnl{l6}
  upon receiving \msg{checkNeighborOp}{side, newNeighbor, $\ell$}:\;
    \lnl{l7}
    \If{(side $=R$)}
    {
        \lnl{l8}
        cmp $ \leftarrow \ <$\;
        \lnl{l9}
        otherSide $ \leftarrow L$\;
    }
    \lnl{l10}
    \Else
    {
        \lnl{l11}
        cmp $ \leftarrow \ >$\;
        \lnl{l12}
        otherSide $ \leftarrow R$\;
    }
    \lnl{l13}
    \If{(\neigh{v}{\textrm{side}}{\ell} $\neq$ newNeighbor)}
    {
        \lnl{l14}
        \If{((\neigh{v}{\textrm{side}}{\ell} $= \nil$) {\bf and}
             ($v$ cmp newNeighbor))}
        {
            \lnl{l15}
            \neigh{v}{\textrm{side}}{\ell} $\leftarrow$ newNeighbor\;
            \lnl{l16}
            \KwSend \msg{checkNeighborOp}{otherSide, $v$, $\ell$} to newNeighbor\;
        }
        \lnl{l17}
        \gIf{((\neigh{v}{\textrm{side}}{\ell} $\neq \nil$) {\bf and}
             (\neigh{v}{\textrm{side}}{\ell} cmp newNeighbor))} 
        {
            \lnl{l18}
            \KwSend \msg{checkNeighborOp}{side, newNeighbor, $\ell$} to
        }
        \lnl{l19}
        \gElse
        {
            \lnl{l20}
            \KwSend \msg{checkNeighborOp}{otherSide, newNeighbor, $\ell$} to \neigh{v}{\textrm{side}}{\ell}\;
            \lnl{l21}
            \KwSend \msg{checkNeighborOp}{otherSide, $v$, $\ell$} to newNeighbor\;
            \lnl{l22}
            \KwSend \msg{checkNeighborOp}{side, \neigh{v}{\textrm{side}}{\ell}, $\ell$} to newNeighbor\;
            \lnl{l23}
            \neigh{v}{\textrm{side}}{\ell} $\leftarrow$ newNeighbor\;
        }
    }
\end{algorithm}


\begin{lemma}
\label{lemma-repair-backpointer}
In the absence of new failures, inserts and deletes, the repair mechanism 
described in Algorithm~\ref{alg-repair-backpointer} repairs any violated 
backpointer constraint without losing existing connectivity.
\end{lemma}
\begin{proof}
We prove that Algorithm~\ref{alg-repair-backpointer} repairs the
violated backpointer constraints for a single node without losing
existing connectivity. We concentrate on the repair of the $R$ 
links as the case for $L$ links is symmetric. 

The violations of Constraint~\ref{c-right-left} for node $v$
at level $\ell$ are as follows:
\begin{enumerate}
  \item $vR_{\ell}=z>v$ but $zL_{\ell}=\nil$: Node $v$ sends 
        \msg{checkNeighborOp}{$L$, $v$, $\ell$} to $z$ (line~\ref{l5}
        of Algorithm~\ref{alg-repair-backpointer}). As $zL_{\ell}=\nil$
        and $z>v$, $z$ sets $zL_{\ell}'=v$ (line~\ref{l15} of
        Algorithm~\ref{alg-repair-backpointer}). Thus after
	Algorithm~\ref{alg-repair-backpointer} finishes,
	$vR_{\ell}'L_{\ell}'=v$, restoring Constraint~\ref{c-right-left}.
  \item $vR_{\ell}=z>v$ but $zL_{\ell}=y>v$: Node $v$ sends 
        \msg{checkNeighborOp}{$L$, $v$, ${\ell}$} to $z$ (line~\ref{l5}
        of Algorithm~\ref{alg-repair-backpointer}). As $zL_{\ell}=y>v$,
        $z$ passes the message on to $y$ (line~\ref{l18} of 
        Algorithm~\ref{alg-repair-backpointer}). As long as this message 
        reaches some node $y>v$ such that $yL_{\ell}>v$, $y$ will pass it
        on to $yL_{\ell}$, until it reaches a node $x>v$ such that
        $xL_{\ell}<v<x$ or $xL_{\ell}=\nil$. Then $x$ sets $xL_{\ell}'=v<x$ 
        (lines~\ref{l15} or~\ref{l23} of Algorithm~\ref{alg-repair-backpointer}).
        Node $x$ also sends \msg{checkNeighborOp}{$R$, $x$, ${\ell}$} to $v$
        (line~\ref{l16} or~\ref{l21} of Algorithm~\ref{alg-repair-backpointer}).
        Upon receiving that message, $v$ sets $vR_{\ell}'=x$.
        Thus after Algorithm~\ref{alg-repair-backpointer} finishes,
	$vR_{\ell}'L_{\ell}'=v$, restoring Constraint~\ref{c-right-left}.
  \item $vR_{\ell}=z>v$ but $zL_{\ell}=u<v$: Node $v$ sends 
        \msg{checkNeighborOp}{$L$, $v$, ${\ell}$} to $z$ (line~\ref{l5}
        of Algorithm~\ref{alg-repair-backpointer}). As $zL_{\ell}=u<v$, $z$
        sets $zL_{\ell}'=v<z$ (line~\ref{l20} of Algorithm~\ref{alg-repair-backpointer}).
        Thus after Algorithm~\ref{alg-repair-backpointer} finishes,
	$vR_{\ell}'L_{\ell}'=v$, restoring Constraint~\ref{c-right-left}.
\end{enumerate}

We prove that the backpointer constraint repair mechanism does not 
lose any existing connectivity in the skip graph, i.e, if a path between 
nodes $v$ and $z$ used to exist before the repair mechanism was initiated,
a path will still exist after the repair mechanism operations finish. 
We consider the case where a node $v$ changes
its successor pointer to a node $z$ and points to another node $y$;
we omit the case where $v$ changes its predecessor pointer as it
is similar to this case. Node $v$ updates its successor pointer 
to some node $z$ at level $\ell$, to point to some other node $y$, 
only when $vR_{\ell}=z>y>v$ (line~\ref{l23} of 
Algorithm~\ref{alg-repair-backpointer}). 
Node $v$ also sends messages to 
(i) $z$ to update its predecessor pointer to point to $y$ 
(line~\ref{l20} of Algorithm~\ref{alg-repair-backpointer}),
(ii) $y$ to update its predecessor pointer to point to $v$
(line~\ref{l21} of Algorithm~\ref{alg-repair-backpointer}) and,
(iii) $y$ to update its successor pointer to point to $z$
(line~\ref{l22} of Algorithm~\ref{alg-repair-backpointer}).
When these messages are delivered, $vR_{\ell}'=y$ and 
$yR_{\ell}'=z$. Thus the path $v$--$z$ is now replaced 
by a longer path $v$--$y$--$z$, and no existing
connectivity is lost.
\end{proof}

It is possible that a node $x$ detects a failed backpointer constraint if it 
checks it while some node $y$ is in the middle of its insert operation. 
Suppose that $xR_{\ell}=z$ but $zL_{\ell}=y$ because $y$ is yet to connect
to $x$. When $x$ sends a \op{checkNeighborOp} message to $z$, it gets passed 
to $y$, which then links to $x$ and asks $x$ to link to it (both through the
repair mechanism and the insert operation). Thus, the repair mechanism generates
additional messages but does not affect the insert operation. In case of a 
delete operation, a node does not delete itself until it has repaired
the links at all the levels so an inconsistent backpointer constraint will 
not be detected during the delete operation.


\subsubsection{Restoring inter-level constraints}
\label{section-repair-successor}

We see how each node periodically checks Constraint~\ref{c-right-successor}; we
omit the details for Constraint~\ref{c-left-successor} as it is a mirror image 
of Constraint~\ref{c-right-successor}. For each level $\ell>0$, each node $x$
sends a message to $xR_{\ell-1}$ to check if $xR_{\ell}=xR_{\ell-1}^k$, for some
$k>0$. Each node that receives the message passes it to the right until one
of the four following cases occur: 

\begin{enumerate}
  \item The message reaches node $a$, $a<xR_{\ell}$ and 
        $m(a)\restrictedto \ell=m(x)\restrictedto\ell$.
  \item The message reaches node $a$, $a<xR_{\ell}$ and $aR_{\ell-1}=\nil$.
  \item The message reaches node $a$, $a=xR_{\ell}$.
        \label{case-all-ok}
  \item The message reaches node $a$, $a>xR_{\ell}$.
\end{enumerate}

In case~\ref{case-all-ok}, Constraint~\ref{c-right-successor} is not
violated and no repair action is violated. We provide a repair mechanism 
for the each of the remaining three cases. The repair mechanism for fixing 
violations of Constraint~\ref{c-left-successor} is symmetric. It may be 
possible to combine the two mechanisms to improve the performance but we 
will treat them separately for simplicity.

In each case, we assume that the link is present at level $\ell$ but absent at
level $\ell-1$. Note that if a node $x$ is linked at level $\ell-1$ but not at
level $\ell$, it can easily traverse the list at level $\ell-1$ to determine
which node to link to at level $\ell$. This process is identical to the
insertion process where a new node uses its neighbors at lower levels
to insert itself at higher levels in the skip graph. 

The violations of Constraint~\ref{c-right-successor} are as follows:
\begin{enumerate}

\item $xR_{\ell} = xR_{\ell-1}^k$, but $\exists a=xR_{\ell-1}^{j}$, 
      \begin{figure}[H]
         \begin{center}
            \input{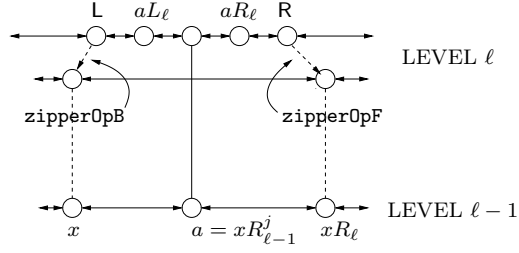}
            \caption{Two-way merge to repair a violated inter-level constraint.}
         \end{center}
      \end{figure}
      The nodes connected to $a$ and $x$ at level $\ell$ have to be merged
      together into one list by sending the following messages:
      \begin{itemize}
        \item Probe level $\ell$ (in list containing $a$)
              to find largest $aR_{\ell}^{k'} = \textsf{R} < xR_{\ell}$.
              Node $a$ starts the probe by sending a message to $aR_{\ell}$.
              Upon reaching node $y$, if $y<xR_{\ell}<yR_{\ell}$, the probe 
              ends with $y = \textsf{R}$. Otherwise the message is passed to
              $yR_{\ell}$.
        \item Send \msg{zipperOpF}{$xR_{\ell}$, $\ell$} to $\textsf{R}$.
        \item Probe level $\ell$ (in list containing $a$)
              to find smallest $aL_{\ell}^{k"} = \textsf{L} > x$.
              Node $a$ starts the probe by sending a message to $aL_{\ell}$.
              Upon reaching node $y$, if $yL_{\ell}<x<y$, the probe ends
              with $y = \textsf{L}$. Otherwise the message is passed to
              $yL_{\ell}$.
        \item Send \msg{zipperOpB}{$x$, $\ell$} to $\textsf{L}$.
      \end{itemize}
     \label{case-present}

\item $xR_{\ell} \neq xR_{\ell-1}^k$ for any $k>0$,  
      and $\exists a < xR_{\ell}, aR_{\ell} = \nil$.
      \label{case-absent-smaller}
      \begin{figure}[H]
          \begin{center}
            \input{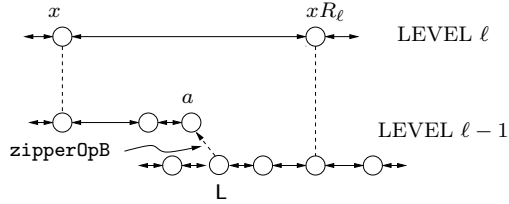}
            \caption{One-way merge to repair a violated inter-level constraint.}
          \end{center}
      \end{figure}
      The nodes connected to $a$ and $xR_{\ell}$ at level $\ell-1$ have to 
      be merged together into one list by sending the following messages:
      \begin{itemize}
          \item Probe level $\ell-1$ (in list containing $xR_{\ell}$)
                to find smallest $xR_{\ell}L_{\ell-1}^{k'}=\textsf{L}>a$. Node
	       	    $xR_{\ell}$ starts the probe by sending a message to 
		        $xR_{\ell}L_{\ell-1}$. Upon reaching node $y$, such that
	     	    $y>a>yL_{\ell-1}$, the probe stops with $y=\textsf{L}$.
		        Otherwise the message is passed on to $yL_{\ell-1}$.
          \item Send \msg{zipperOpB}{$a$, $\ell-1$} to \textsf{L}.
      \end{itemize}
      \label{case-single}

\item $xR_{\ell} \neq xR_{\ell-1}^{j}$ for any $j>0$,
      and $\exists a = xR_{\ell-1}^k > xR_{\ell}$.
      \label{case-absent-bigger}
      \begin{figure}[H]
          \begin{center}
            \input{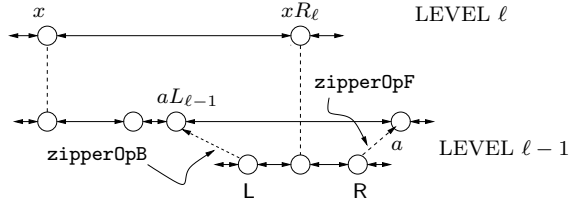}
            \caption{Two-way merge to repair a violated inter-level constraint.}
          \end{center}
      \end{figure}
      The nodes connected to $a$ and $xR_{\ell}$ at level $\ell-1$ have to be merged
      together into one list by sending the following messages\footnote{Details
      of the \op{zipperOp} messages are given in Algorithms~\ref{alg-zipperOpB}
      and~\ref{alg-zipperOpF}.}:
      \begin{itemize}
        \item Probe level $\ell-1$ (in list containing $xR_{\ell}$)
              to find largest $xR_{\ell}R_{\ell-1}^{k'} = \textsf{R} < a$.
              Node $xR_{\ell}$ starts the probe by sending a message to 
              $xR_{\ell}R_{\ell-1}$.
	      Upon reaching node $y$, if $y<a<yR_{\ell-1}$, the probe ends with 
	      $y=\textsf{R}$. Otherwise the message is passed to $yR_{\ell-1}$.
        \item Send \msg{zipperOpF}{$a$, $\ell-1$} to $\textsf{R}$.
        \item Probe level $\ell-1$ (in list containing $xR_{\ell}$)
              to find smallest $xR_{\ell}L_{\ell-1}^{k"} = \textsf{L} > aL_{\ell-1}$.
              In this case, the probe proceeds along the predecessors of
              $xR_{\ell}$ at level ${\ell}-1$ till it reaches node $y$ such that
	          $y=\textsf{L}>aL_{\ell-1}>yL_{\ell-1}$.
        \item Send \msg{zipperOpB}{$aL_{\ell-1}$, $\ell-1$} to $\textsf{L}$.
      \end{itemize}
      \label{case-double}

\end{enumerate}


\begin{algorithm}[htp]
\caption{\op{zipperOpB} for node $v$
\label{alg-zipperOpB}}
    \lnl{zb1}
    upon receiving \msg{zipperOpB}{$x$, $\ell$}:\;
      \lnl{zb2}
      \uIf{\neigh{v}{L}{\ell} $> x$}
      {
         \lnl{zb3}
         \KwSend \msg{zipperOpB}{$x$, $\ell$} to \neigh{v}{L}{\ell}\;
      }
      \lnl{zb4}
      \Else
      {
        \lnl{zb5}
        tmp $=$ \neigh{v}{L}{\ell}\;
        \lnl{zb6}
        \neigh{v}{L}{\ell} $=x$\;
        \lnl{zb7}
        \KwSend \msg{updateOp}{$R$, $v$, $\ell$} to $x$\;
        \lnl{zb8}
        \If{tmp $\neq \nil$}
        {
           \lnl{zb9}
           \KwSend \msg{zipperOpB}{tmp, $\ell$} to $x$\;
        }
      }
\end{algorithm}


\begin{algorithm}[htp]
\caption{\op{zipperOpF} for node $v$
\label{alg-zipperOpF}}
    \lnl{zf1}
    upon receiving \msg{zipperOpF}{$x$, $\ell$}:\;
      \lnl{zf2}
      \If{\neigh{v}{R}{\ell} $< x$}
      {
         \lnl{zf3}
         \KwSend \msg{zipperOpF}{$x$, $\ell$} to \neigh{v}{R}{\ell}\;
      }
      \lnl{zf4}
      \Else
      {
        \lnl{zf5}
        tmp $=$ \neigh{v}{R}{\ell}\;
        \lnl{zf6}
        \neigh{v}{R}{\ell}$=x$\;
        \lnl{zf7}
        \KwSend \msg{updateOp}{$L$, $v$, $\ell$} to $x$\;
        \lnl{zf8}
        \If{tmp $\neq \nil$}
        {
           \lnl{zf9}
           \KwSend \msg{zipperOpF}{tmp, $\ell$} to $x$\;
        }
      }
\end{algorithm}


\begin{figure}[htp]
    \centerline{\input{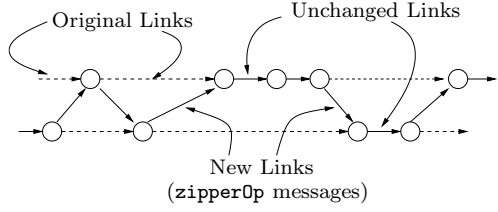}}
    \caption{\op{zipperOp} operation to merge nodes on the same level.}
  \label{figure-zipperOp}
\end{figure}


\begin{lemma}
\label{lemma-repair-successor}
In the absence of new failures, inserts and deletes, the repair mechanism 
described in Section~\ref{section-repair-successor} repairs any violated 
inter-level constraint without losing existing connectivity.
\end{lemma}
\begin{proof}
The algorithm initiates repair for all the possible violations of the
inter-level constraints of a node as given above. It only remains to 
be proved that the \op{zipperOp} messages merge two sorted lists at
a given level into a single sorted list, without losing existing
connectivity. We concentrate on the \op{zipperOpF} messages as the 
\op{zipperOpB} messages are symmetric.

To prove that the repair mechanism merges two sorted lists into a 
single sorted one, we first see that a node $v$ always receives a 
\op{zipperOpF} message to 
link to a node greater than itself. The initial \op{zipperOpF}
messages sent are as follows: (i) In Case~\ref{case-absent-bigger},
\textsf{R} receives a \op{zipperOpF} message to link to $a>$ \textsf{R},
and (ii) in Case~\ref{case-present}, \textsf{R} receives a \op{zipperOpF}
message to link to $xR_{\ell}>$ \textsf{R}. When a node $v$ receives
a \op{zipperOpF} message to link to $x$, if $vR_{\ell}<x$, it sends
the message to $vR_{\ell}$ to link to $x$ (line~\ref{zf3}
of Algorithm~\ref{alg-zipperOpF}). Otherwise, it updates
$vR_{\ell}'=x$ (line~\ref{zf6} of Algorithm~\ref{alg-zipperOpF}), and it 
sends a \op{zipperOpF} message to $x<vR_{\ell}$ to link to $vR_{\ell}$
(line~\ref{zf9} of Algorithm~\ref{alg-zipperOpF}). In both cases,
the \op{zipperOpF} message reaches a node that has to link to a node
greater than itself. Also, each node $v$ only links to a new node $x$ 
if it is smaller than the current successor of $v$, so 
$v<vR_{\ell}=x<vR_{\ell}$. Thus the two sorted lists get merged 
into a single sorted list, until one of the lists terminates.

We also prove that the inter-level constraint repair mechanism does not 
lose any existing connectivity in the skip graph, i.e, if a path between 
nodes $v$ and $z$ used to exist before the repair mechanism was initiated,
a path will still exist after the repair mechanism operations finish. A link 
change occurs only when $v$ receives a \op{zipperOpF} message to link to 
$x<vR_{\ell}=z$. Node $v$ sends a \op{zipperOpF} message to $x$ to
link to $z$. Upon receipt of that message, $x$ either 
sets to $xR_{\ell}'=z$, or it sends passes the message to 
$xR_{\ell}<z$. In the first case, the path between $v$
and $z$ is replaced by a new longer path $v$-$x$-$z$.
In the latter case, the \op{zipperOp} message passes through 
several nodes $xR_{\ell}, xR_{\ell}^2,\ldots$, until it reaches a 
node $y$ such that $y<z<yR_{\ell}$ and $y$ sets 
$yR_{\ell}'=z$. Then the path $v$--$z$ is replaced by a longer 
path $v$--$x$--$xR_{\ell}$--$xR_{\ell}^2$--$\ldots$--$y$--$z$,
and no existing connectivity is lost.
\end{proof}

It is possible that a node $x$ detects a failed inter-level constraint if it 
checks it while some node $y$ is in the middle of its insert or delete operation. 
Node $x$ will detect a failed constraint at level $\ell$ if $y$ has inserted
itself at level $\ell-1$ and not at level $\ell$. The probe message along
level $\ell-1$ will reach $y$ which can then inform $x$ that it is yet
to complete its insert operation, and thus terminate the repair mechanism.


\subsection{Proof of correctness}
\label{section-put-together}

In this section we prove that the repair mechanism given in 
Algorithm~\ref{alg-repair-backpointer} and 
Section~\ref{section-repair-successor} repairs a defective
a defective skip graph. 

We prove that the repair mechanism repairs a defective skip graph
by showing that it repairs level $0$ after some finite interval of 
time, and then uses the links at level $0$ to restore the links
at higher levels.
Lemma~\ref{lemma-two-level-R} and Corollary~\ref{coro-all-level-R}
show that if there exists a path between two nodes that consists
entirely of pointers in any one direction ($L$ or $R$), then the
repair mechanism ensures that after some finite interval of time,
there is a path between those two nodes in the same direction at 
level $0$. For their proofs, we consider only the case for the $R$
links as the case for the $L$ links is symmetric. This result is
further extended in Lemma~\ref{lemma-collapse} which shows that 
as long as there is path between two nodes, irrespective of the
directions of the edges in the path, there will be a path in
both directions between the two nodes at level $0$.
Corollary~\ref{coro-sorted-level-0} shows that this leads to a single, 
sorted, doubly-linked list at level $0$ as in a defectless skip graph. 
Finally, Lemma~\ref{lemma-sorted-level-l} and Theorem~\ref{theorem-repair}
show how the list at level $0$ is used to create lists at higher
levels, to eventually give a defectless skip graph.

\begin{lemma}
\label{lemma-two-level-R}
  \item 1. Suppose we have $y_0 R_{\ell_1} R_{\ell_2} \ldots R_{\ell_r}=y_r$, 
        $y_0<y_r$, for some $r$, and for each $1 \leq i \leq r$, $\ell \leq 
        \ell_i \leq \ell+1$. Then after some finite interval of time, 
        $y_0 R_{\ell}^k = y_r$, for some $k$.
  \item 2. Suppose we have $y_0 L_{\ell_1} L_{\ell_2} \ldots L_{\ell_r}=y_r$, 
        $y_0>y_r$, for some $r$, and for each $1 \leq i \leq r$, $\ell \leq 
        \ell_i \leq \ell+1$. Then after some finite interval of time, 
        $y_0 L_{\ell}^k = y_r$, for some $k$.
\end{lemma}
\begin{proof}
Let $y_0 R_{\ell_1} R_{\ell_2} \ldots R_{\ell_i} =y_i$. Then there exists
a link between each $y_{i-1}$ and $y_{i}$ at level $\ell$ or $\ell+1$. 
For each $R_{\ell_i}=R_{\ell+1}$, as $y_{i-1}R_{\ell+1}=y_{i}$, the inter-level 
repair mechanism given in Section~\ref{section-repair-successor}
ensures that after some finite interval of time, $y_{i-1}R_{\ell}^{k_i}=y_{i}$,
for some $k_i$ (Lemma~\ref{lemma-repair-successor}). 
We then have $y_0 R_{\ell_1} R_{\ell_2} \ldots R_{\ell_r}
=y_0 R_{\ell}^{k_1} R_{\ell}^{k_2} \ldots R_{\ell}^{k_r}=y_{r}$, where
$k_i=1$ if $R_{\ell_i}=R_{\ell}$. Thus we get $y_0 R_{\ell}^k=y_{r}$, 
where $k=k_1+k_2+\ldots+k_r$. 
\end{proof}


\begin{corollary}
\label{coro-all-level-R}
   \item 1. Suppose we have $y_0 R_{\ell_1} R_{\ell_2} \ldots R_{\ell_r}= y_r$, $y_0<y_r$,
         for some $r$, and for each $i$, $1\leq i \leq r$, $\ell_i\geq 0$.
         Then after some finite interval of time, $y_0 R_0^k = y_r$, for some $k$.
   \item 2. Suppose we have $y_0 L_{\ell_1} L_{\ell_2} \ldots L_{\ell_r}= y_r$, $y_0>y_r$,
         for some $r$, and for each $i$, $1\leq i \leq r$, $\ell_i\geq 0$.
         Then after some finite interval of time, $y_0 L_0^k = y_r$, for some $k$.
\end{corollary}
\begin{proof}
Let $y_0 R_{\ell_1} R_{\ell_2} \ldots R_{\ell_i} =y_i$. Then for each $y_{i-1}$,
$y_{i-1}R_{\ell_i}=y_i$. By Lemma~\ref{lemma-two-level-R}, after some finite 
interval of time, $y_{i-1}R_{\ell_i-1}^{k_{i, \ell_i-1}}=y_i$, for some 
$k_{i, \ell_i-1}$. By repeatedly applying Lemma~\ref{lemma-two-level-R}, after 
some finite interval of time, we get $y_{i-1}R_0^{k_{i, 0}}=y_i$, for some
$k_{i, 0}$. So we get $y_0 R_0^{k_{1, 0}} R_0^{k_{2, 0}} \ldots
R_0^{k_{r, 0}}=y_r$. Thus, $y_0 R_0^k=y_r$, where 
$k=k_{1, 0}+k_{2, 0}+\ldots+k_{r, 0}$.
\end{proof}


\begin{lemma}
\label{lemma-collapse}
Suppose we have $y_0 E_{\ell_1} E_{\ell_2} \ldots E_{\ell_r}= y_r$, for some $r$,
$y_0<y_r$, $E \in \{L, R\}$, and for each $1 \leq i \leq r$, $\ell_i\geq 0$. 
Then after some finite interval of time, $y_0 R_0^k = y_r$ and
$y_r L_0^k = y_0$ for some $k$.
\end{lemma}
\begin{proof}
Let $y_0 E_{\ell_1} E_{\ell_2} \ldots E_{\ell_i} = y_i$.
For each $y_i$, $y_{i-1} E_{\ell_i} = y_i$. By Corollary~\ref{coro-all-level-R},
after some finite interval of time, $y_{i-1} E_0^{k_i}=y_i$ for some $k_i$.
So we have $y_0 E_0^{k_1} E_0^{k_2} \ldots E_0^{k_r}=y_r$. 
If any of the $y_i$'s are
not distinct, then we can eliminate the path between two consecutive 
occurrences of $y_i$. So we can replace a path of the form 
$y_i E_0^{k_i} \ldots y_j E_0^{k_j}$, where $y_i=y_j$, with 
$y_i E_0^{k_j}$. Thus we have a path consisting of $L_0$ and 
$R_0$ edges starting at $y_0$ and terminating at $y_r$ which 
consists only of unique nodes. 

For each node, after some finite interval of time, the backpointer 
constraint repair mechanism given in Algorithm~\ref{alg-repair-backpointer} 
will repair any violated backpointer constraints without losing 
existing connectivity (Lemma~\ref{lemma-repair-backpointer}). 
So Constraints~\ref{c-right-left} and~\ref{c-left-right} are repaired for 
all the nodes in the path, and as proved in Theorem~\ref{theorem-maintain}, 
Constraints~\ref{c-right-bigger} and~\ref{c-left-smaller} are 
always maintained. As proved in Theorem~\ref{theorem-iff-skip-graph},
with Constraints~\ref{c-right-bigger} through~\ref{c-left-right}
satisfied for all the nodes in the path, we get a sorted, doubly-linked list 
of the nodes. Thus, $y_0 R_0^k=y_r$ and $y_r L_0^k=y_0$, for some $k$.
\end{proof}


\begin{corollary}
\label{coro-sorted-level-0}
After some finite interval of time, all nodes in the same
connected component of a skip graph are linked together in
a single, sorted, doubly-linked list at level $0$.
\end{corollary}
\begin{proof}
Lemma~\ref{lemma-collapse} shows that any two connected nodes $x$
and $y$ are in the same sorted, doubly-linked list at level $0$ 
after some finite interval of time. Any other node $z$ in the 
same connected component is also connected to both $x$ and $y$, 
so it has to be in the same list at level $0$. Thus, all the 
nodes in the same connected component are in a single list
at level $0$.
\end{proof}


Given a set $S$, let $M_{\ell}(S)$ be the set of all membership vector prefixes 
of length $\ell$ represented by the nodes in $S$,
i.e., $M_{\ell}(S) = \{w\;|\;\exists x\in S, m(x)\restrictedto \ell=w\}$.

\begin{lemma}
\label{lemma-sorted-level-l}
Suppose we have all nodes in the same connected component $C$ of a skip 
graph linked together in $|M_{\ell}(C)|$ sorted, doubly-linked lists at 
level $\ell$, one for each $w \in M_{\ell}(C)$. Then after some finite 
interval of time, we get $|M_{\ell+1}(C)|$ sorted, doubly-linked lists at 
level $\ell+1$, one for each $w \in M_{\ell+1}(C)$.
\end{lemma}
\begin{proof}
Consider a single list $\mathbb{L}$ at level $\ell$, and let node
$x\in\mathbb{L}$. As explained in the successor 
constraint repair mechanism given in Section~\ref{section-repair-successor},
if $xL_{\ell}, xR_{\ell} \notin \{\nil\}$, $x$ uses its neighbors at level 
$\ell$, to find its neighbors at level $\ell+1$. As all the lists at level 
$\ell$ are sorted and doubly-linked, $x$ can find $xR_{\ell+1}$ and $xL_{\ell+1}$ 
in $O(2|\Sigma|)$ time, just like in an insert operation
(Algorithm~\ref{alg-insert-new-node}). When all the nodes of 
list $\mathbb{L}$ determine their neighbors at level $\ell+1$ after some finite 
interval of time,  we get the lists from the nodes of $\mathbb{L}$ at level $\ell+1$, 
one for each $w \in M_{\ell+1}(\mathbb{L})$. This is identical to the insert operation 
and as proved in Lemma~\ref{lemma-insert-preserve}, all the lists thus
created are sorted and doubly-linked, and only consist of nodes that have the
matching membership vector prefix of length $\ell+1$. Thus, considering all 
the $|M_{\ell}(C)|$ lists at level $\ell$, after some finite interval of time, 
we get $|M_{\ell+1}(C)|$ sorted, doubly-linked lists at level $\ell+1$, one for 
each $w \in M_{\ell+1}(C)$.
\end{proof}


\begin{theorem}
\label{theorem-repair}
The repair mechanism given in Section~\ref{section-repair} repairs a
defective skip graph to give a defectless skip graph after some finite 
interval of time.
\end{theorem}
\begin{proof}
Corollary~\ref{coro-sorted-level-0} shows that after some finite interval
of time, the repair mechanism links all the nodes in the same connected
component in a single, sorted, doubly-linked list at level $0$. Further,
there are only finitely many levels in a skip graph. Using 
Lemma~\ref{lemma-sorted-level-l} inductively, with 
Corollary~\ref{coro-sorted-level-0} as the base case, we can show that we
get sorted, doubly-linked lists, which contain all the nodes with the
matching non-empty membership vector prefixes, at all levels of the skip graph. 
Thus, the repair mechanism repairs a defective skip graph to give a
defectless skip graph after some finite interval of time.
\end{proof}


\section{Congestion}
\label{section-congestion}

In addition to fault tolerance, a skip graph provides a limited
form of congestion control, by smoothing out hot spots caused by popular
search targets.  The guarantees that a skip graph  makes in this case
are similar to the guarantees made for survivability. Just as a node's
continued connectivity depends on the survival of its neighbors, its
message load depends on the popularity of its neighbors as search
targets.
However, we can show that this effect drops off rapidly with distance;
nodes that are far away from a popular target in the bottom-level
list of a skip graph get little increased message load on average.

We give two versions of this result.  The first version, 
given in Section~\ref{section-congestion-single-search}, shows that
the probability that a particular search uses a node between the
source and target drops off inversely with the distance from the node
to the target.
This fact is not necessarily reassuring to heavily-loaded nodes.
Since the probability averages over all choices of membership vectors,
it may be that some particularly unlucky node finds itself with a
membership vector that puts it on nearly
every search path to some very popular target.
The second version, given in 
Section~\ref{section-congestion-multiple-searches}, shows that our
average-case bounds hold with high probability. While it is still
possible that a spectacularly unlucky node is hit by most searches,
such a situation only occurs for very low-probability choices of
membership vectors.  It follows that most skip graphs alleviate
congestion well. For our results, we consider skip graphs with
alphabet $\{0,1\}$.


\subsection{Average congestion for a single search}
\label{section-congestion-single-search}

Our argument that the average congestion is inversely proportional to
distance is based on the observation that a node only appears on a
search path in a skip list $S$ if it is among the tallest nodes
between itself and the target.  We will need a small technical lemma
that counts the expected number of such tallest nodes.
Consider a set-valued Markov process $A_0 \supseteq A_1 \supseteq
A_2 \ldots$ where $A_0$ is some nonempty initial set
and each element of $A_t$
appears in $A_{t+1}$ with independent probability $\frac{1}{2}$.
Let $\tau$ be the largest index for which $A_\tau$ is not empty.
We will now show that $\E[|A_\tau|]$ is small regardless of the
size of the initial set $A_0$.

\pagebreak
\begin{lemma}
\label{lemma-pruning}
Let $A_0, A_1, \ldots, A_\tau$ be defined as above.
Then
$\E[|A_\tau|] < 2$.
\end{lemma}
\begin{proof}
The bound on $\E[|A_\tau|]$ will follow from a surprising
connection between $\E[|A_\tau|]$ and $\Pr[|A_\tau|=1]$.
We begin by obtaining a recurrence for $\Pr[|A_\tau|]=1$.

Let $P(n) = \Pr[|A_\tau| = 1 : |A_0| = n]$.
Clearly $P(0) = 0$ and
$P(1) = 1 \ge \frac{2}{3}$.
For larger $n$, summing over all $k = |A_1|$ gives
$P(n) = 2^{-n} \sum_{k = 0}^{n} {n \choose k} P(k)$,
which can be rewritten as
$P(n) = \frac{1}{2^n-1} \sum_{k=1}^{n-1} {n \choose k} P(k)$.
The solution to this recurrence goes
asymptotically to
$\frac{1}{2 \ln 2} \pm 10^{-5} \approx 0.7213\cdots$~\cite[Theorem
7.9]{SedgewickF1996};
however, we will use the much simpler property that when $n\ge 2$, 
$P(n)=\Pr[|A_\tau| = 1]$ is the probability of an event that does 
not always occur, and is thus less than $1$.

Let $E(n) = \E[|A_\tau| : |A_0| = n]$.
Then
$E(n) = 2^{-n} \left(n + \sum_{k = 1}^{n} {n \choose k} E(k)\right)$,
and eliminating the $E(n)$ term on the right-hand side gives
$E(n) = \frac{1}{2^n-1} \left(n + \sum_{k=1}^{n-1} {n \choose k}
E(k)\right)$.

For $n=1$, $E(1)=1$ by definition.
Recall that $P(1)$ is also equal to $1$.
We will now show that $E(n) = 2P(n)$ for all $n > 1$.
Suppose that $E(k) = 2P(k)$ for $1 < k < n$.
Let $n=2$, then $E(k)=2P(k)$ for all $1<k<n$ (an empty set
of $k$). Then,
\begin{eqnarray*}
E(n)
&=& \frac{1}{2^n-1} \left(n + \sum_{k=1}^{n-1} {n \choose k}
E(k)\right)
\\
&=&
\frac{1}{2^n-1} \left(n + {n \choose 1} E(1) + 2 \sum_{k=2}^{n-1} {n
\choose k} P(k)\right)
\\
&=&
\frac{1}{2^n-1} \left(2 {n \choose 1} P(1) + 2 \sum_{k=2}^{n-1} {n
\choose k} P(k)\right)
\\
&=&
2\cdot \frac{1}{2^n-1} \sum_{k=1}^{n-1} {n
\choose k} P(k)
\\
&=&
2P(n).
\end{eqnarray*}

Since $P(n) < 1$ for $n > 1$, we immediately get $E(n) < 2$ for all
$n$.  For large $n$ this is an overestimate: given the
asymptotic behavior of $P(n)$, $E(n)$ approaches
$\frac{1}{\ln 2} \pm 2\times 10^{-5} \approx 1.4427\cdots$.
But it is close enough for our purposes.
\end{proof}

\begin{theorem}
\label{theorem-congestion-balancing-single-search}
Let $S$ be a skip graph with alphabet $\{0,1\}$,
and consider a search from $s$ to $t$ in $S$.
Let $u$ be a node with $s < u < t$ in the key ordering
(the case $s > u > t$ is symmetric),
and let $d$ be the distance from $u$
to $t$, defined as the number of nodes $v$ with $u < v \le t$.
Then the probability that a search from $s$ to $t$ passes through $u$
is less than $\frac{2}{d+1}$.
\end{theorem}
\begin{proof}
Let $S_{m(s)}$ be the skip list restriction of $s$ whose existence is
shown by Lemma~\ref{lemma-restriction}. From Lemma~\ref{lemma-search-time},
we know that searches in $S$ follow searches in $S_{m(s)}$.
Observe that for $u$ to appear in the search path from $s$ to $t$ in
$S_{m(s)}$ there must be no node $v$ with $u < v \le t$ whose height in 
$S_{m(s)}$ is higher than $u$'s.
It follows that $u$ can appear in the search path only if it is among
the tallest nodes in the interval $[u,t]$, i.e., if
$|m(s) \wedge m(u)| \ge |m(s) \wedge m(v)|$ for all $v$ with $u < v
\le t$.  Recall that $m(s) \wedge m(v)$ is the common prefix
(possibly empty) of $m(s)$ and $m(v)$.

There are $d+1$ nodes in this interval.
By symmetry, if there are $k$ tallest nodes then the probability that
$u$ is among them is $\frac{k}{d+1}$. Let T be the random variable
representing the set of tallest nodes in the interval. Then:

\begin{eqnarray*}
\Pr[u \in T]
&=&
\sum_{k=1}^{d+1} \Pr[|T| = k] \frac{k}{d+1}
\\
&=&
\frac{\E[|T|]}{d+1}.
\end{eqnarray*}

What is the expected size of $T$?  All $d+1$ nodes have height at
least $0$, and in general each node with height at least $k$ has
height at least $k+1$ with independent probability $\frac{1}{2}$.  The
set $T$ consists of the nodes that are left at the last level
before all nodes vanish.  It is thus equal to $A_\tau$ in the process
defined in Lemma~\ref{lemma-pruning}, and we have $\E[|T|] < 2$
and thus $\Pr[u \in T] < \frac{2}{d+1}$.
\end{proof}

\begin{figure}[htp]
  \begin{center}
  \epsfig{figure=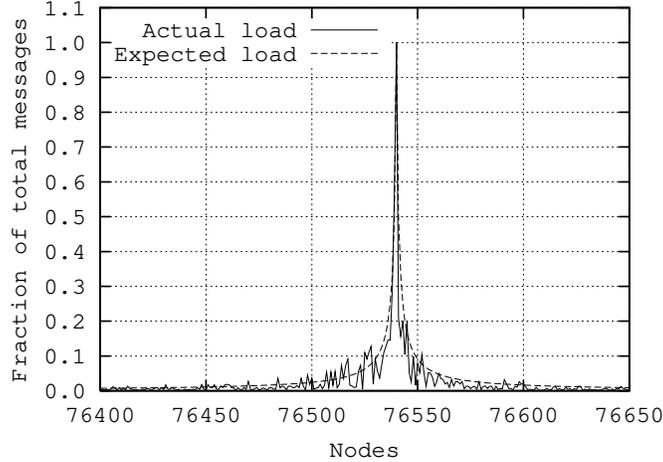, width=255pt}
  \caption[Congestion in the nodes of a skip graph.]
  {Actual and expected congestion in a skip graph with $131072$ nodes
  with the target=$76539$. Messages were delivered from each node to the
  target and the actual number of messages through each node was measured.
  The bound on the expected congestion is computed using 
  Theorem~\ref{theorem-congestion-balancing-single-search}. Note that this
  bound may overestimate the actual expected congestion.}
\label{fig-congestion}
\end{center}
\end{figure}

For comparison, experimental data for the congestion in a skip
graph with $131072$ nodes, together with the theoretical average
predicted by Theorem~\ref{theorem-congestion-balancing-single-search},
is shown in Figure~\ref{fig-congestion}.

\subsection{Distribution of the average congestion}
\label{section-congestion-multiple-searches}

Theorem~\ref{theorem-congestion-balancing-single-search} is of small
consolation to some node that draws a probability $\frac{2}{d+1}$ straw and
participates in every search.  Fortunately, such disasters do not happen
often.  Define the \buzz{average
congestion} $L_{tu}$ imposed by a search for $t$ on a node $u$
as the probability that an $s-t$ search hits $u$
conditioned on the membership vectors of all nodes in the interval
$[u,t]$, where $s < u < t$, or, equivalently, 
$s > u > t$.\footnote{It
is immediate from the proof of
Theorem~\ref{theorem-congestion-balancing-single-search} that $L_{tu}$ does
not depend on the choice of $s$.}
Note that since the conditioning does not include the membership
vector of $s$, the definition in effect assumes that $m(s)$ is chosen
randomly.
This approximates the situation in a fixed skip graph
where a particular target $t$ is used for many searches that may hit
$u$, but the sources of these searches are chosen randomly from the
other nodes in the graph.

Theorem~\ref{theorem-congestion-balancing-single-search} implies that the
expected value of $L_{tu}$ is no more than $\frac{2}{d+1}$.  In the
following theorem, we show the distribution of $L_{tu}$ declines
exponentially beyond this point.

\begin{theorem}
\label{theorem-congestion-balancing-multiple-searches}
Let $S$ be a skip graph with alphabet $\{0,1\}$.
Fix nodes $t$ and $u$, where $u < t$ and
$|\{v : u < v \le t\}| = d$.
Then for any integer ${\ell} \ge 0$,
$\Pr[L_{tu} > 2^{-{\ell}}] \le 2 e^{-2^{-{\ell}}d}$.
\end{theorem}
\begin{proof}
Let $V = \{ v : u < v \le t \}$ and let $m(V) = \{ m(v) : v \in V \}$.
As in the proof of Theorem~\ref{theorem-congestion-balancing-single-search},
we will use the fact that $u$ is on
the path from $s$ to $t$ if and only if $u$'s height in $S_{m(s)}$ is not
exceeded by the height of any node $v$ in $V$.

To simplify the notation, let us assume without loss of generality that $m(u)$
is the all-zero vector.  Then the height of $u$ in $S_{m(s)}$ is equal to
the length of the initial prefix of zeroes in $m(s)$, and $u$ has
height at least ${\ell}$ with probability $2^{-{\ell}}$.  Whether this is enough
to raise it to the level of the tallest nodes in $V$ will depend on
what membership vectors appear in $m(V)$.

Let $m(s) = 0^i1\cdots$.  Then $u$ has height exactly $i$, and is hit
by an $s-t$ search unless there is some $v \in V$ with $m(v) =
0^{i}1\cdots$.  We will argue that when $d = |V|$ is sufficiently large,
then there is a high probability that all initial prefixes $0^{i}1$
appear in $m(V)$ for $i < {\ell}$.
In this case, $u$ can only appear in the $s-t$ path if its height is at
least ${\ell}$, which occurs with probability only $2^{-{\ell}}$.
So if $0^{i}1$ appears as a prefix of some $m(v)$ for all $i < {\ell}$,
then
$L_{tu} \le 2^{-{\ell}}$.
Conversely, if $L_{tu} > 2^{-{\ell}}$, then $0^{i}1$ does {\em not} appear as a
prefix of some $m(v)$ for some $i < {\ell}$.

Now let us calculate the probability that not all such prefixes appear in
$m(V)$.  We are going to show that this probability is at most
$2e^{-2^{-{\ell}}d}$, and so we need to consider only the case where
$e^{-2^{-{\ell}}d} \le \frac{1}{2}$; this bound is used in
steps~(\ref{eqn-ltu-two-j-to-j-plus-one})
and~(\ref{eqn-ltu-geometric}) below.  We have:
\begin{eqnarray}
\Pr[L_{tu} > 2^{-{\ell}}]
&\le&
\Pr[\neg\left(\forall i < {\ell}: \exists v \in V: 0^i1 \prefix
m(v)\right)]
\nonumber\\
&=&
\Pr[\exists i < {\ell}: \forall v \in V: 0^i1 \not\prefix m(v)]
\nonumber\\
&\le&
\sum_{i=0}^{{\ell}-1} \Pr[\forall v \in V: 0^i1 \not\prefix m(v)]
\nonumber\\
&=&
\sum_{i=0}^{{\ell}-1} \left(1-2^{-i-1}\right)^{d}
\nonumber\\
&\le&
\sum_{i=0}^{{\ell}-1} e^{-2^{-i-1}d}
\nonumber\\
&=&
\sum_{j=0}^{{\ell}-1} e^{-2^{-{\ell}+j}d}
\nonumber\\
&=&
\sum_{j=0}^{{\ell}-1} \left(e^{-2^{-{\ell}}d}\right)^{2^j}
\nonumber\\
&\le&
\sum_{j=0}^{{\ell}-1} \left(e^{-2^{-{\ell}}d}\right)^{j+1}
\label{eqn-ltu-two-j-to-j-plus-one}
\\
&\le&
\sum_{j=0}^{\infty} \left(e^{-2^{-{\ell}}d}\right)^{j+1}
\nonumber\\
&=&
\frac{e^{-2^{-{\ell}}d}}{1-e^{-2^{-{\ell}}d}}
\nonumber\\
&\le&
2e^{-2^{-{\ell}}d}.
\label{eqn-ltu-geometric}
\end{eqnarray}
\end{proof}


\section{Related Work}
\label{section-related}

SkipNet is a system very similar to skip graphs that was independently 
developed by Harvey~\etal~\cite{HarveyJSTW2003}. 
SkipNet builds a trie of circular, singly-linked skip lists 
to link the {\em machines} in the system. The machines names are sorted 
using the domain in which they are located (for example {\tt www.yale.edu}). 
In addition to the pointers between all the machines in all the domains
that are structured like a skip graph, within each individual domain, 
the machines are also linked using a DHT, and the resources are uniformly 
distributed over all the machines using hashing. A search consists of two 
stages: First, the search locates the domain in which a resource lies by
using a search operation similar to a skip graph. Second, once the search
reaches some machine inside a particular domain, it uses greedy routing as 
in DHTs to locate the resource within that domain. SkipNet has been 
successfully implemented, and this shows that a skip-graph-like structure
can be used to build real systems.

The SkipNet design ensures \buzz{path locality} i.e., the traffic within 
a domain traverses other nodes only within the same domain. Further,
each domain gets to hosts it own data which provides \buzz{content 
locality} and inherent security. Finally, using the hybrid storage 
and search scheme provides \buzz{constrained load balancing} within 
a given domain.
However, as the name of the data item includes the domain in which it
is located, transparent remapping of resources to other domains is 
not possible, thus giving a very limited form of load balancing.
Another drawback of this design is that it does not give full-fledged
spatial locality. For example, if the resources are document files,
sorting according to the domain on which they are served 
gives no advantage in searching for related files compared to DHTs.

Zhang~\etal~\cite{ZhangSZ2003} and Awerbuch~\etal~\cite{AwerbuchS2003}
have both independently suggested designs for peer-to-peer systems
using separate data structures for resources and the machines that
store them. The main idea is to build a data structure $D$ over the
resources, which are distributed uniformly among all the 
machines using hashing, and to build a separate DHT
over all the machines in the system. Each resource maintains the keys 
of its neighboring resources in $D$, and each machine maintains the 
addresses of its neighboring machines as per the DHT network. To access 
a neighbor $b$ of resource $a$, $a$ initiates a DHT search for the hash 
value of $b$. One pointer access in $D$ is converted to a search operation 
in the DHT, so if any operation in $D$ takes time $t$, the same operation 
takes $O(t\log m)$ time with $m$ machines in this hybrid system.
Zhang~\etal~\cite{ZhangSZ2003} focus on implementing a tree of  the
resources, in which each node in the tree is responsible for some fixed 
range of the keyspace that its parent is responsible for.
Awerbuch~\etal~\cite{AwerbuchS2003} propose building a skip graph of the 
resources on top of the machines in the DHT.

This design approach is interesting because it allows building any
data structure using the resources, while providing uniform load
balancing. In particular, both these systems support complex 
queries as in skip graphs, and uniform load balancing as in DHTs. We 
believe that distributing the resources uniformly among all the nodes 
(as described in Section~\ref{section-implementation}) will also have 
the same properties as these two approaches.

However, the Awerbuch~\etal~$\;$approach and our uniform resource distribution
approach suffer from the same problems of high storage requirements and 
high volume of repair mechanism message traffic as a skip graph.  
With $m$ machines and $n$ resources in the system, in the 
Awerbuch~\etal~$\;$approach, each machine has to store $O(\log m)$ pointers 
(for the DHT links) 
and $O(\log n)$ keys for {\em each} resource that it hosts (for the 
skip graph pointers). Further, the repair mechanism has to repair 
the broken DHT links as well as inconsistent skip graph keys. Finally, 
the search performance is degraded to $O(\log^2 m)$ compared to 
$O(\log m)$ in DHTs and $O(\log n)$ in skip graphs. In comparison,
in our approach of uniformly resource distribution, each machine
has to store $O(\log n)$ pointers for each resource that it hosts, repair is 
required only for the skip graph links, and the search time is $O(\log n)$ as 
in skip graphs. 

In the Zhang~\etal~$\;$approach, each machine has to store $k$ keys
for the $k$ children of each tree node that it hosts, and $O(\log m)$
pointers for the DHT links. Repair involves fixing broken tree keys
as well as broken DHT links. This scheme suffers from the other
problems of tree data structures such as increased traffic on the nodes
higher up in the tree, and vulnerability to failures of these nodes.
Further, unlike skip graphs, it require a priori knowledge about the 
keyspace in order to assign specific ranges to the tree nodes. Thus,
it is an open problem to design a system that efficiently supports
both uniform load balancing and complex queries.


\section{Conclusion}
\label{section-conclusion}

We have defined a new data structure, the skip graph,
for distributed data stores that has several desirable properties.
Constructing, inserting new nodes into, and searching in a skip graph
can be done using simple and straightforward algorithms. 
Skip graphs are highly resilient, tolerating a large fraction of
failed nodes without losing connectivity. Using the repair mechanism, 
disruptions to the data structure can be repaired in the absence of 
additional faults. Skip graphs also support range
queries which allows, for example, searching for a copy of a resource
near a particular location by using the location as low-order field
in the key and clustering of nodes with similar keys.

As explained in Section~\ref{section-related}, one issue that remains to 
be addressed is the large number of pointers per machine in the system. 
It would be interesting to design a peer-to-peer system that maintains 
fewer pointers per machine and yet supports spatial locality. Also, skip 
graphs do not exploit network locality (such as or latency along 
transmission paths) in location
of resources and it would be interesting to study performance benefits
in that direction, perhaps by using multi-dimensional skip graphs.
As with other overlay networks, it would be interesting to
see how the network performs in the presence of Byzantine failures.
Finally, it would be useful to develop a more efficient repair mechanism 
and a self-stabilization mechanism to repair defective skip graphs.


\section{Acknowledgments}
We would like to thank Arvind Krishnamurthy for helpful discussions.


\bibliographystyle{alpha}
\bibliography{paper}


\end{document}